\documentclass[prb,twocolumn,amsmath,amssymb,groupedaddress]{revtex4} 
\usepackage{graphicx}
\usepackage{enumitem}
\usepackage{bbold}
\usepackage[colorlinks,bookmarks=false,citecolor=red,linkcolor=blue,urlcolor=blue]{hyperref}

\newcommand{\bea}{\begin{eqnarray}}
\newcommand{\eea}{\end{eqnarray}}

\usepackage{diagbox}
\allowdisplaybreaks

\begin{document}

\title{Collective excitations in vortices and vortex lattices}
\author{Bhargava B. A.}
\affiliation{The Institute of Mathematical Sciences, HBNI, C I T Campus, Chennai 600 113, India}
\affiliation{IFW Dresden and W\"urzburg-Dresden Cluster of Excellence ct.qmat, Helmholtzstr. 20, 01069 Dresden, Germany}
\author{R. Ganesh}
\affiliation{The Institute of Mathematical Sciences, HBNI, C I T Campus, Chennai 600 113, India}
\email{ganesh@imsc.res.in}
\date{\today}

\begin{abstract}
Emergent lattices at mesoscopic length scales have evoked interest in several recent contexts, e.g., in crystalline arrangements of skyrmions. It is a challenging task to determine their collective excitations as the unit cells are large and can contain complex textures. We address this issue in the oldest known example of an emergent mesoscopic lattice, the vortex lattice in a superfluid. 
We show that a tight binding approach can successfully describe collective modes. 
We begin with a single isolated vortex in a two dimensional system. With suitable pinning mechanims, the low energy excitations take the form of localized pairing fluctuations. 
They can be viewed as wave-like disturbances that propagate around the vortex centre, in the azimuthal direction. In particular, the lowest energy excitations are gyrotropic and breathing modes. The former corresponds to circular motion of the eye of the vortex, while the latter represents an oscillation between sharper and broader vortex profiles. These modes are `bosonic' analogues of atomic orbitals with the vortex profile playing the role of the nuclear potential. 
Moving to a sparse vortex lattice with well separated vortices, the single-vortex excitations provide a convenient basis for finding normal modes. We derive an analogue of Bloch's theorem in this context, revealing a non-trivial phase contribution arising from the orbital field. We set up a tight binding prescription and use it to explicitly determine the band structure of excitations about a square vortex lattice. 
Our results can be tested in ultracold atomic gases with synthetic magnetic fields.
\end{abstract}
\pacs{}
                                 
\keywords{}
\maketitle

\tableofcontents

\section{Introduction}
Crystal structures and band theory are fundamental building blocks of condensed matter physics. The notion of electronic bands emerges as a consequence of spatial periodicity, as brought out by Bloch's theorem. 
While traditional examples of crystals are solid materials with nanoscale lattice periods, new examples have emerged with periods that are larger in order of magnitude. A recent example is twisted bilayer graphene wherein a rotation between two graphene layers leads to a Moir\'e pattern with a lattice constant of the order of $\sim$100 Angstroms\cite{Bistritzer2011,Cao2018,Cao2018b}. The large mesoscopic unit cell poses difficulties in determining the band structure. From a tight binding perspective, it is difficult to identify a basis of `atomic' orbitals that span the bands of interest. At a deeper level, it can be difficult to determine the correct structure of the tight binding model that respects all symmetries and is free of `obstructions'\cite{Po2019}. As with twisted bilayer graphene, these considerations may give rise to fruitful insights in other contexts as well.

Apart from crystalline arrangements of atoms, mesoscopic spatial periodicity also occurs in `emergent' lattices.  
This is typically seen in ordered phases where topological defects arrange themselves in a regular pattern. A recent example is the phenomenon of skyrmion crystallization where magnetic order occurs with a large unit cell\cite{Liu2016,Finocchio2016,Fert2017}. Other examples include crystals of visons\cite{Zhang2019}, $\mathbb{Z}_2$ vortices\cite{Rousochatzakis2016,Seabrook2019} and merons\cite{KarmakarPRB2017}. The oldest and best known example is the vortex lattice phase in type-II superconductors. Since Abrikosov's proposal\cite{Abrikosov1957} in 1957, vortex lattices have been studied theoretically and experimentally in several contexts: in conventional and unconventional superconductors, liquid He\cite{Sonin1987} and ultracold atomic gases\cite{Fetter2009}. In this article, we describe the low energy excitations of a vortex lattice, focussing on how they are constrained by lattice periodicity. In the process, we also characterize the excitations about a single isolated vortex in the presence of pinning mechanisms.  
Our focus is on collective excitations, i.e., fluctuations of the order parameter field that propagate coherently with a well-defined dispersion relation. 
 While we place our discussion within the context of superfluids with short-ranged interactions, our approach can be generalized to a wide array of textured phases. 

\subsection*{Structure of this paper}
The rest of this article is structured as follows. In Sec.~\ref{sec.framework}, we introduce the time-dependent Gross Pitaevskii equation, define normal modes and set up the formalism in broad terms. In Sec.~\ref{sec.isolated}, we discuss collective modes about a single vortex that is centred at the origin, drawing parallels with atomic orbitals. We discuss the physical character of low lying excitations, classifying them into gyrotropic and breathing modes. We also discuss 
the necessity of pinning 
to obtain a consistent description of low energy excitations. 
We move to vortex lattices in Sec.~\ref{sec.VL} and discuss their transformations under lattice translations. We show that the underlying gauge structure manifests as a non-trivial phase accrued with each translation. In Sec.~\ref{sec.bloch}, we discuss collective modes in a vortex lattice. We derive an analogue of Bloch's theorem and define the associated notion of Wannier functions. We use these ideas to implement a tight binding prescription in Sec.~\ref{sec.tb}. We present an approach based on linear combinations of atomic orbitals, where the atomic orbitals are collective modes about a single vortex. As an illustration, we determine the band structure of collective modes in a square vortex lattice. We conclude with a summary and discussion in Sec.~\ref{sec.summary}.

\section{Framework}
\label{sec.framework}
Landau-Ginzburg theory provides a suitable theoretical framework for studying superfluidity and superconductivity. Order parameter configurations in equilibrium satisfy
\bea
\Big[\frac{1}{2m^*}\big( -i\hbar\boldsymbol{\nabla} +2e\textbf{A} (\mathbf{r})\big)^{2} + a + 2b \vert \Psi  (\mathbf{r}) \vert ^{2} \Big]\Psi  (\mathbf{r}) = 0,
\label{eq.psi_eq}
\eea 
where $a$ and $b$ are system parameters. In the ordered phase, $a$ and $b$ take negative and positive values respectively. 
The order parameter $\Psi (\mathbf{r})$ is a complex-valued field. In a superconductor, it carries charge and couples to the external vector potential via the minimal coupling scheme. In this article, we focus on neutral superfluids where the coupling to the vector potential can be `simulated', e.g., by rotation of the condensate\cite{Fetter2009,Schweikhard2004,Lieb2006,Lin2009}. We will assume a uniform static magnetic field, $\mathbf{B}(\mathbf{r})=B\hat{z}$ as is appropriate for synthetic magnetic fields. We choose the symmetric gauge, so that the vector potential is given by $\textbf{A}(\mathbf{r}) = \frac{B}{2}\{ x\hat{y} - y\hat{x} \}$. 

In order to study fluctuations, we begin with $\Psi_0(\mathbf{r})$, a given solution to Eq.~\ref{eq.psi_eq}. We consider a small fluctuation that depends on space and time with $\Psi(\mathbf{r},t) \equiv \Psi_{0} (\mathbf{r}) + \eta (\mathbf{r},t)$. Its time evolution is determined by the time-dependent Gross-Pitaevskii equation\cite{PethickSmith},
\bea
\nonumber \Big[\frac{1}{2m^*}\big\{ -i\hbar\boldsymbol{\nabla} &+& 2e\textbf{A}(\mathbf{r})\big\}^{2} + a 
+ 2b\vert\Psi_0(\mathbf{r})\vert ^{2}\Big] \eta(\mathbf{r},t)  \\
&+& b\Psi_0^{2}(\mathbf{r})\eta^{*}  (\mathbf{r},t)= i\hbar\frac{\partial \eta (\mathbf{r},t)}{\partial t}.~~
\label{eq.SE}
\eea 
This equation is directly analogous to the time-dependent Schr\"odinger equation. 
It can be brought into a time-independent form by defining
\bea
\eta(\mathbf{r},t) = u(\mathbf{r})e^{i\frac{\epsilon}{\hbar}t} - v^{*}(\mathbf{r})e^{-i\frac{\epsilon}{\hbar}t}.
\label{eq.etaform}
\eea
With this substitution, Eq.~\ref{eq.SE} reduces to a matrix eigenvalue equation,
\bea
\left(
\begin{array}{cc}
-\hat{\mathcal{H}}_{-2e} &  b\Psi_{0}^{2} (\mathbf{r}) \\
 -b\Psi_{0}^{*2}(\mathbf{r}) & \hat{\mathcal{H}}_{2e}  
\end{array}
\right) \left(
\begin{array}{c}
u(\mathbf{r}) \\
v(\mathbf{r})
\end{array} 
\right)= \epsilon \left(
\begin{array}{c}
u(\mathbf{r}) \\
v(\mathbf{r})
\end{array} 
\right),
\label{eq.uveqs}
\eea
where $\hat{\mathcal{H}}_q =  \frac{1}{2m^*}\big( -i\hbar\boldsymbol{\nabla} - q\textbf{A}(\mathbf{r})\big)^{2} + a + 2b\vert \Psi_{0} (\mathbf{r}) \vert^{2} $.
 As Eq.~\ref{eq.SE} is linear, its most general solution can be obtained by linearly superposing all independent `normal modes'. The normal modes are, in fact, the eigenstates found in Eq.~\ref{eq.uveqs}. Note that the `Hamiltonian' matrix in Eq.~\ref{eq.uveqs} is non-Hermitian. This is a common feature of collective fluctuations about ordered phases, e.g., of spin waves in a Heisenberg antiferromagnet. We take the eigenvectors to satisfy an orthonormality relation given by
 $\int d^2 r \big\{u_i^*(\mathbf{r}) u_j (\mathbf{r}) - v_i^*(\mathbf{r}) v_j (\mathbf{r}) \big\} = -\mathrm{sign}(\epsilon_i) \delta_{ij}$.  
With this choice of normalization, we can identify $\vert \epsilon \vert$ as the energy of the normal mode. This can be seen as follows. We consider an excited state given by $\Psi_\alpha(\mathbf{r},t) = \Psi_{0}(\mathbf{r}) + \alpha ~\eta_{NM} (\mathbf{r},t)$. Here, $\alpha$ is a small real number, representing the amplitude of the fluctuation. 
The function $\eta_{NM}(\mathbf{r},t)$ encodes a normal mode, as given in Eq.~\ref{eq.etaform} with $u(\mathbf{r})$, $v(\mathbf{r})$ satisfying Eq.~\ref{eq.uveqs}. 
 The free energy cost of this fluctuation can be found by substituting $\Psi_\alpha (\mathbf{r},t)$ in the free energy functional\cite{PethickSmith}. With the above normalization condition, the free energy cost comes out to be $\alpha ^2 \vert \epsilon \vert$. We thus interpret $\vert \epsilon\vert$ as the normal mode energy.

Below, we will first find solutions for $u(\mathbf{r})$ and $v(\mathbf{r})$ where the reference state, $\Psi_0 (\mathbf{r})$, corresponds to an isolated vortex. We will then use these solutions to construct a tight binding prescription for fluctuations about a vortex lattice.

\section{Excitations about an isolated vortex }
\label{sec.isolated}
We consider a solution to Eq.~\ref{eq.psi_eq} of the form $\Psi_{vortex} (\mathbf{r})= \Delta(r) e^{-i\theta}$, where $r$ and $\theta$ are the usual polar coordinates. The amplitude depends purely on the radial coordinate, vanishing at the origin and asymptotically approaching the amplitude of the uniform solution, i.e., $\Delta(0)=0$ and $\Delta(\infty) = \Delta_0=\sqrt{-a/b}$, where $a$ and $b$ are the (Landau-Ginzburg) parameters in Eq.~\ref{eq.psi_eq}.
The amplitude profile is well approximated by $\Delta(r) \approx \Delta_0 \tanh(\nu r /\xi)$, where $\xi$ represents the coherence length of the superconductor and $\nu\approx 1/\sqrt{2}$ is a numerical pre-factor\cite{PethickSmith,Verhelst2017}.

\subsection{Expansion in Landau levels}
To study fluctuations about a single isolated vortex, we consider Eq.~\ref{eq.uveqs} with $\Psi_0 (\mathbf{r})=\Psi_{vortex}(\mathbf{r})$. Each diagonal entry in Eq.~\ref{eq.uveqs} resembles the Hamiltonian of a particle in a magnetic field subject to a rotationally symmetric potential.
This suggests that $u(\mathbf{r})$ and $v(\mathbf{r})$ can be expanded in the basis of Landau level wavefunctions (in the symmetric gauge). We take
\bea
\nonumber u(\mathbf{r}) &\equiv& \sum_{n=0}^{\infty} \sum_{m=-n}^{\infty} u_{n,m} \big(\psi^{LL}_{n,m} (\mathbf{r})\big)^*, \\
v(\mathbf{r}) &\equiv& \sum_{n=0}^{\infty} \sum_{m=-n}^{\infty} v_{n,m} \psi^{LL}_{n,m} (\mathbf{r}).
\label{eq.uv_LL_expansion}
\eea
The basis states represent stationary states in the problem of a free particle in a magnetic field. They can be written as $\psi^{LL}_{n,m} (\mathbf{r}) = e^{im\theta} f_{n,m}^{LL}(r)$. The amplitude, $f_{n,m}^{LL}(r)$, is localized near the origin, decaying exponentially over a length scale set by the magnetic length. The $m$ index can be viewed as the angular momentum, as it determines the winding of the phase.
We choose $\big(\psi^{LL}_{n,m} (\mathbf{r})\big)^*$ and $\psi^{LL}_{n,m} (\mathbf{r})$ as basis elements for $u(\mathbf{r})$ and $v(\mathbf{r})$ respectively, as this leads to a convenient description for low energy excitations below.

In order to calculate spectra, it is necessary to restrict the Landau level expansion by imposing a cutoff. For concreteness, we only keep the lowest Landau level in the discussion below, i.e., we keep only $n=0$. In Appendix~\ref{app.spectrum}, we show that the results do not change qualitatively upon including more Landau levels.   
In Sec.~\ref{ssec.landau} below, we argue that the Landau level cutoff serves as a proxy for pinning mechanisms in the problem. In the remainder of this article, in the interest of simplicity, we restrict the Landau level expansion to $n=0$ and drop the $n$ index, e.g., expressing $u_{n,m}$ as $u_m$.

In addition, we narrow our focus to the three lowest angular momentum values, i.e., $m=0,1,2$. These sectors host the lowest energy excitations, see Appendix~\ref{app.spectrum} for a detailed discussion.

To calculate fluctuation modes about an isolated vortex, we plug in the expansions into Eq.~\ref{eq.uveqs}. The orthogonality properties of Landau level wavefunctions lead to a simple $2\times 2$ matrix equation,
\bea
\left(
\begin{array}{cc}
-a_{m} & b_{m,2-m} \\
-b_{2-m,m} & a_{2-m}
\end{array}\right) 
\left(\begin{array}{c}
u_{m} \\ 
v_{2-m}
\end{array}\right) = \epsilon \left(\begin{array}{c}
u_{m} \\ 
v_{2-m}
\end{array}\right).
\label{eq.uv_eigeqn}
\eea
Explicit expressions for the entries $a_m$, $b_{m,2-m}$, etc. are given in Appendix~\ref{app.spectrum}. 
These entries contain matrix elements of Landau levels with the profile of an isolated vortex. 
Their values depend upon the ratio of two length scales. The first is $\xi$, the coherence length or the size of the vortex. The second is the magnetic length, $\ell_B$. The latter enters in the amplitudes of Landau level wavefunctions, i.e., in $f_{n,m}^{LL}(r)$. 
We evaluate the entries in Eq.~\ref{eq.uv_eigeqn} numerically for any given value of $(\xi/\ell_B)$. We find the same qualitative results for eigenvectors and eigenvalues for various choices of the ratio. We discuss qualitative features of our solutions below. Details about the numerical results are given in Appendix~\ref{app.spectrum}.

\subsection{Analogy with atomic orbitals}
We discuss the physical nature of the low-lying excitations about an isolated vortex, calculated using the formalism described above. We first draw an analogy between low energy modes and `atomic orbitals' known from solving the Schr\"odinger equation in a central Coulomb potential. The latter are localized near the nucleus with the Bohr radius serving as a characteristic length scale. Likewise, the excitations about an isolated vortex are centred around the vortex core. The magnetic length serves as the analogue of the Bohr radius. This property is inherited from the basis states of the lowest Landau level.
This can be seen from the explicit form of the order parameter when a normal mode is excited,

\bea
\nonumber  \Psi(\mathbf{r},t) = \Psi_{vortex}(\mathbf{r}) &+& \alpha \big\{ u_m \big(\psi_{0,m}^{LL} (\mathbf{r}) \big)^* e^{\frac{i}{\hbar}\epsilon t} \\
  &-& v_{2-m}^* \big(\psi_{0,2-m}^{LL} (\mathbf{r}) \big)^* e^{-\frac{i}{\hbar}\epsilon t} \big\},
\label{eq.psim0}
\eea
where $\alpha$ is the fluctuation amplitude. 
We have two Landau level wavefunctions corresponding to $(n,m)=(0,m)$ and $(0,2-m)$. These are both localized near the origin and decay exponentially at distances greater than the magnetic length.

Atomic orbitals can be viewed as waves in the space around the nucleus. 
Such an interpretation can also be given to fluctuation normal modes about an isolated vortex.
We first note that the latter represent modulations that are periodic in time. For instance, in Eq.~\ref{eq.psim0}, the fluctuation is periodic with time period $T=h/\epsilon$. To interpret it in terms of waves, we take out a common factor of $e^{-i\theta}$ in Eq.~\ref{eq.psim0} above,
\bea
\nonumber \Psi(\mathbf{r},t) =e^{-i\theta} \Big[ \Delta(r)  &+& \alpha \big( u_m f_{0,m}^{LL} (r) e^{\frac{i}{\hbar}\epsilon t - i(m-1)\theta} \\
&+& v_{2-m}^* f_{0,2-m}^{LL}(r) e^{-\frac{i}{\hbar}\epsilon t+i(m-1)\theta} \big)\Big].~~~~
\label{eq.psifluc}
\eea
The expression within the brackets corresponds to the order parameter in the co-moving frame, where the phase winding of the reference vortex has been removed. This is reflected in the static part, $\Delta(r) \sim \Delta_0 \tanh(\nu r/\xi)$, that contains only the amplitude of an isolated vortex. 
The remaining terms constitute waves that propagate in the azimuthal ($\hat{\theta}$) direction. We use this picture to describe the lowest energy modes below.

\begin{figure*}
\includegraphics[width=2.05\columnwidth]{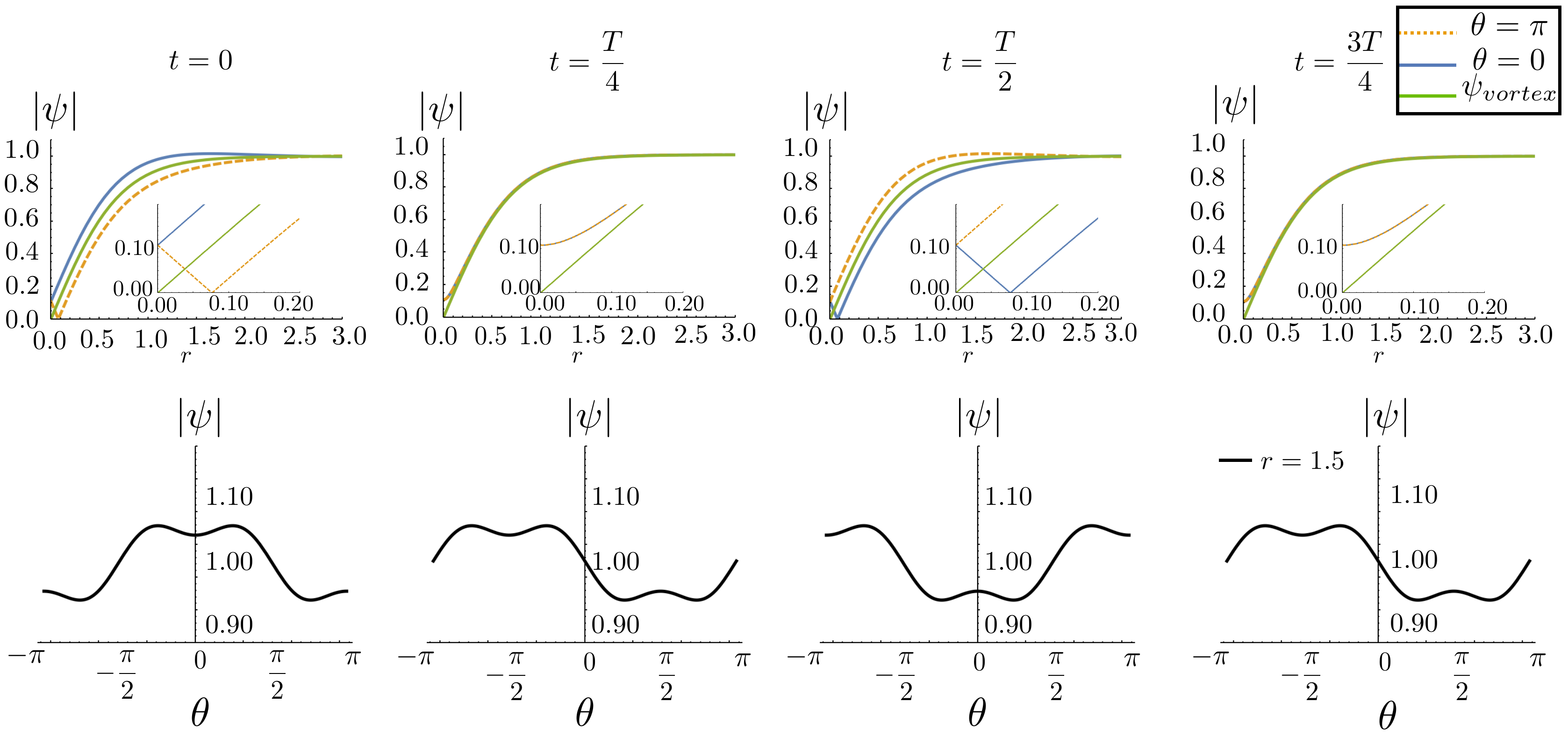}
\caption{Gyrotropic normal mode: 
We depict the order parameter when a normal mode with $m=0$ is excited. The profile is calculated assuming $\xi/\ell_B = 1/2$ and $\alpha = 0.25$. We show the amplitude profile at four different times, with $T$ denoting the time period. Panels on top: Order parameter amplitude vs. $r$. The insets show the same profiles over a narrow range in the vicinity of the origin. The green line represents $\psi_{vortex}(r)$, the reference solution or the profile of the static vortex. 
Bottom panels: Amplitude vs. $\theta$ at a fixed $r$. In all panels, the amplitude is plotted in units of $\Delta_{0}$.}
\label{fig.fluc_amp}
\end{figure*}

\begin{figure}
\includegraphics[width=\columnwidth]{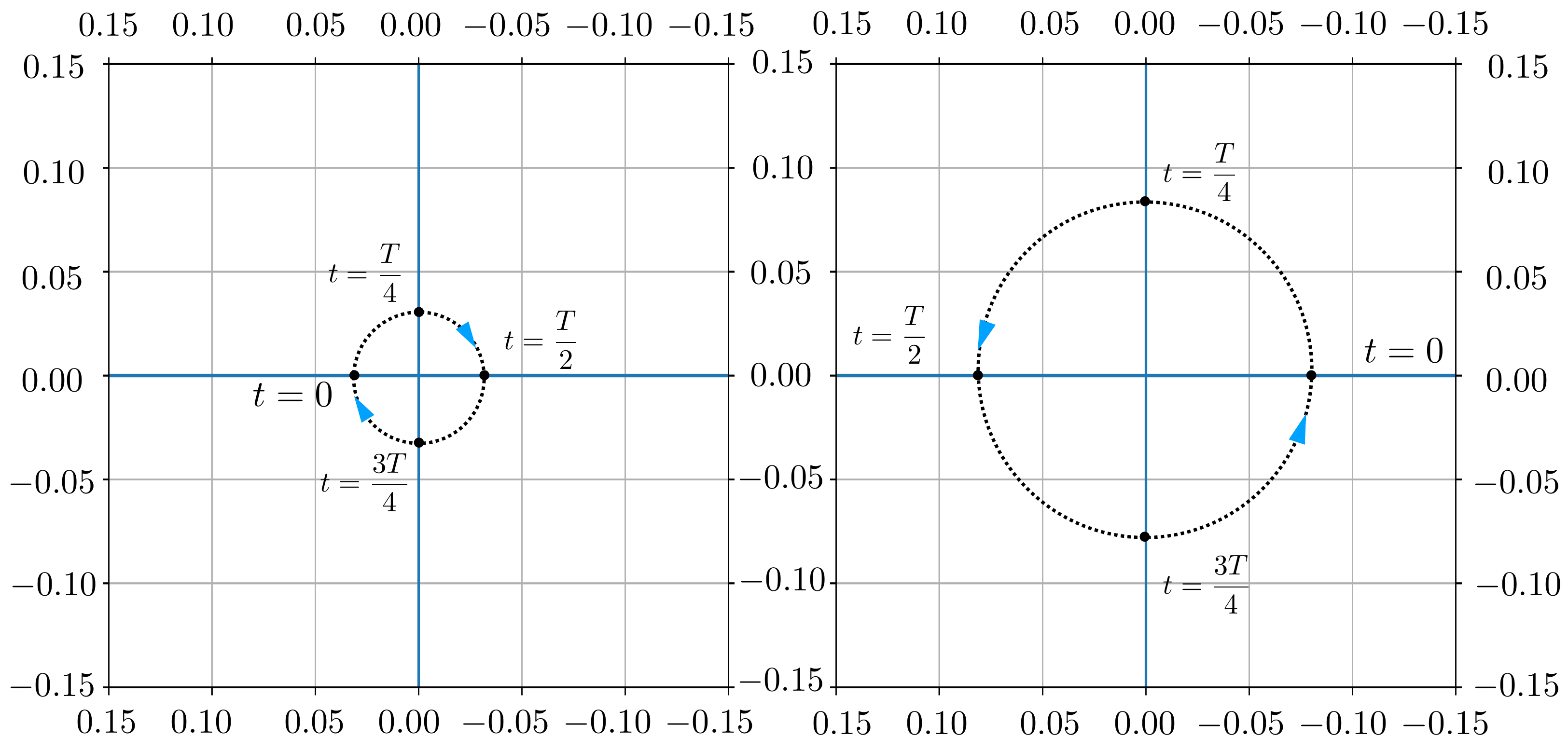}
\caption{
Motion of the eye of the vortex when a gyrotropic mode is excited. The two panels correspond to the two independent gyrotropic modes. In both, the eye executes circular motion. The position of the eye at specific times is indicated, with $T$ representing the time period. The positions have been calculated assuming $\xi/\ell_B = 1/2$ and $\alpha = 0.25$.  }
\label{fig.gyro}
\end{figure}

Below, we focus on specific low energy modes and their character. We designate them as `gyrotropic' and `breathing' modes, following terminology developed in the context of magnetic vortices and skyrmions\cite{Onose2012,Buttner2015,Kravchuk2018,Kravchuk2019}.
  
\subsection{Gyrotropic modes}
The winding of the phase around a vortex forces its amplitude to vanish at its centre. In the unperturbed vortex, this singles out one point as the `eye' of the vortex. We designate an excitation as a gyrotropic mode if it leads to a time-dependent displacement of the eye. In particular, we find modes where the eye undergoes periodic circular motion. In our formalism, the reference vortex has its eye at the origin. A fluctuation can shift the eye only if it has non-zero contributions from the $m=0$ states in the Landau level expansion. The basis states with $m\neq0$ vanish at the origin and are therefore incapable of displacing the eye.

We obtain gyrotropic modes as eigenstates of Eq.~\ref{eq.uv_eigeqn} when $m=0$. There are two such modes with different energies, corresponding to two independent eigenvectors in Eq.~\ref{eq.uv_eigeqn}. The same physical modes are obtained by choosing $m=2$ in Eq.~\ref{eq.uv_eigeqn}, as this simply leads to a rearrangement of the matrix.  
One of the gyrotropic modes is depicted in Fig.~\ref{fig.fluc_amp}. The top panels show the order parameter amplitude vs. $r$, the radial coordinate. The profile is shown for two different polar angles, $\theta=0$ and $\pi$. We see that the fluctuation is localized, with no deviation from the static vortex profile at large distances. In the vicinity of the origin, we see the zero of the amplitude shifting with time. 
The bottom panels show the order parameter amplitude vs. $\theta$, at a fixed radial distance. This reveals an oscillatory pattern. With increasing time, this oscillation moves coherently in the direction of increasing $\theta$.  

The gyrotropic character is shown in Fig.~\ref{fig.gyro}. This shows the position of the eye of the vortex as a function of time. As there are two independent $m=0$ modes, the motion of eye of the vortex is shown separately for each excitation. We see that the modes correspond to circular motion of the eye in clockwise and counter-clockwise directions. The radius of the circles is directly proportional to the fluctuation amplitude, given by $\alpha$ in Eq.~\ref{eq.psifluc}.

\subsection{Breathing modes}
In analogy with skyrmionic systems, we designate an excitation as a breathing mode if it preserves rotational symmetry in the order parameter amplitude. Such a mode arises as a solution to Eq.~\ref{eq.uv_eigeqn} when $m$ is chosen to be unity. This choice leads to two separate modes, obtained as independent eigenvectors of Eq.~\ref{eq.uv_eigeqn}. When either of these modes is excited, the order parameter takes the form,
\bea
\Psi(\mathbf{r},t) =e^{-i\theta} \Big[ \Delta(r)  + \alpha f_{0,1}^{LL} (r) \big( u_1 e^{\frac{i}{\hbar}\epsilon t} 
- v_1^* e^{-\frac{i}{\hbar}\epsilon t} \big)\Big].
\label{eq.psifluc_1}
\eea
The polar angle enters solely in the overall phase. As a result, the order parameter amplitude is independent of $\theta$ at all times, i.e., it remains rotationally symmetric.
 
\begin{figure*}
\includegraphics[width=2.05\columnwidth]{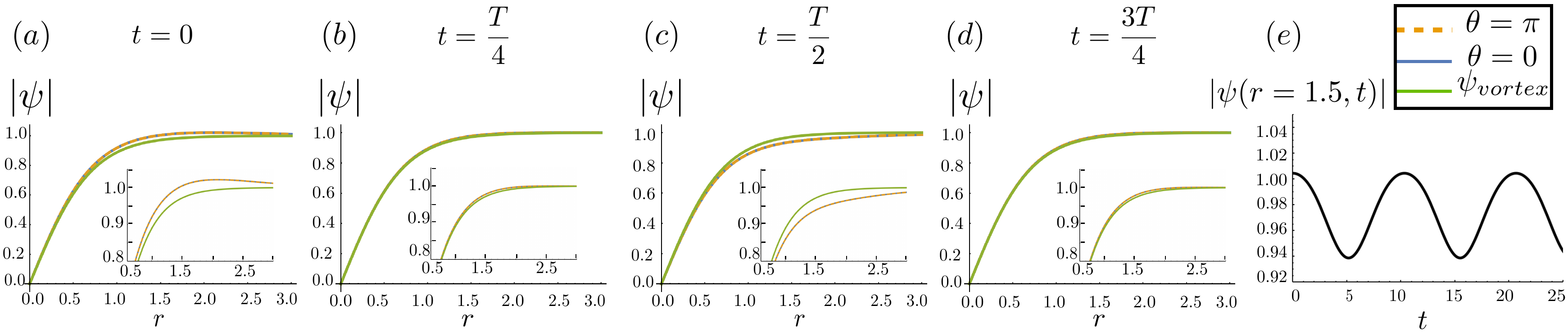}

\caption{Breathing mode: We depict the order parameter amplitude when a normal mode with $m=1$ is excited. The profile is calculated assuming $\xi/\ell_B = 1/2$ and $\alpha = 0.25$. (a-d) Amplitude vs. $r$ at different times, with $T$ representing the time period. The three curves depict the the reference vortex solution, the amplitude at $\theta=0$ and that at $\theta=\pi$. The latter two are identical. The insets show the same data in a narrower range, where the deviation from the reference solution can be clearly seen. (e) Amplitude vs. $t$ at a fixed value of position. In all panels, the amplitude is plotted in units of $\Delta_{0}$.} 
\label{fig.fluc_amp_1}
\end{figure*}

Fig.~\ref{fig.fluc_amp_1} shows the order parameter profile in the presence of a breathing mode fluctuation. In panels (a-d), we see the order parameter amplitude vs. $r$, the radial coordinate. The profile is shown for two fixed polar angles, $\theta=0$ and $\pi$. 

At a given $\theta$, we see that the profile oscillates in time about the reference vortex profile. The oscillations at all $\theta$'s are in phase, preserving rotational symmetry about the origin. 
In panel (e), we plot the amplitude at a fixed position as a function of time, revealing clear oscillations. Put together, the breathing mode corresponds to rotationally symmetric oscillations in the amplitude. This can be crudely visualized as an oscillation in the coherence length, with the vortex oscillating between sharper and broader profiles. This can be seen by examining the profiles in panels (a-d) near the origin, where the order parameter increases linearly from zero. The slope oscillates in time, revealing a `breathing' fluctuation. 

\subsection{Role of pinning}
\label{ssec.landau}
To summarize the discussion so far, we have derived the normal modes for fluctuations about an isolated vortex. We use a truncated expansion in the basis of Landau levels that gives localized modes. We have focussed on gyrotropic ($m=0,2$) and breathing ($m=1$) modes, elucidating their character. These constitute the lowest energy excitations about an isolated vortex (see Appendix~\ref{app.spectrum}). We now show that this approach is consistent only if pinning mechanisms are present in the problem.

We first point out two symmetries in the problem and discuss their consequences for the normal mode spectrum:

(a) Spatial translations: The vortex solution to Eq.~\ref{eq.psi_eq}, $\Psi_{vortex} (\mathbf{r}) = \Delta_0 (r) e^{-i\theta}$, is centred at the origin. In physical terms, this point does not have any special significance. However, within our gauge choice, it is the point where the vector potential vanishes. The zero of the vector potential as well as the centre of the vortex can be shifted elsewhere by a gauge transformation. To see this, we consider Eq.~\ref{eq.psi_eq} with the position vector shifted by an arbitrary vector, $\mathbf{r}_s$,
\bea
\nonumber \Big[\frac{1}{2m^*}\big( &-&i\hbar\boldsymbol{\nabla} +2e\textbf{A} (\mathbf{r}-\mathbf{r}_s)\big)^{2} + a \\
&+& 2b \vert \Psi_{vortex}  (\mathbf{r}-\mathbf{r}_s) \vert ^{2} \Big]\Psi_{vortex}  (\mathbf{r}-\mathbf{r}_s) = 0,~~~
\label{eq.psi_eq_shift}
\eea 
This can be viewed as a gauge transformation with
\bea
\textbf{A} (\mathbf{r}-\mathbf{r}_{s}) = \textbf{A} (\mathbf{r}) + \boldsymbol{\nabla} \lambda_{s} (\mathbf{r}),
\label{eq.Atrans}
\eea
where $\lambda_{s} (\mathbf{r}) =  \frac{B}{2} \{\hat{z} \cdot \mathbf{r} \times \mathbf{r}_{s}\}$ is a scalar function. This relation follows from the form of the vector potential in the symmetric gauge. By undoing the gauge transformation in Eq.~\ref{eq.psi_eq_shift}, we find a new solution to Eq.~\ref{eq.psi_eq}, given by 
\bea
\Psi_{vortex,\mathbf{r}_s}(\mathbf{r}) \equiv e^{2ie \lambda_{s}(\mathbf{r})/\hbar}\Psi_{vortex}  (\mathbf{r}-\mathbf{r}_{s}).
\eea
This solution represents a single vortex centred at $\mathbf{r}_s$. It also has the same energy as the unshifted vortex solution centred at the origin. 

These arguments indicate translational symmetry upto a gauge-derived phase. We may expect this property to result in a Goldstone-like mode with zero energy cost. 
Physically, such a mode would correspond to an infinitesimal shift in the vortex position. 
We expect this mode to appear in the $m=0$ sector, as all Landau level states with $m\neq0$ vanish at the origin and are incapable of shifting the eye of the vortex. 
However, this corresponds to a large and long-ranged change in the order parameter configuration. 
The fluctuation, $(\Psi_{vortex,\mathbf{r}_s} (\mathbf{r}) -\Psi_{vortex} (\mathbf{r}))$, remains significant even as $r\rightarrow \infty$ on account of the gauge-derived phase. 
This implies that, in order to access the physics of the Goldstone mode, we must have a fluctuation mode that extends over all space. This cannot be accessed within our operational framework as our basis states (Landau levels) are all exponentially localized. In other words, this physics will only appear when we keep all orders in the Landau level expansion.

In Appendix~\ref{app.spectrum}, we show that the $m=0$ sector indeed remains gapped even as more Landau levels are included. This indicates that Goldstone-like mode only appears when the Landau level cutoff is taken to infinity.

(b) Global phase change: The Landau Ginzburg equation in Eq.~\ref{eq.psi_eq} is invariant under a global phase change. We may add a phase to the isolated vortex solution to give $\Psi_{vortex,\gamma} (\mathbf{r} ) = \Delta_0 (r) e^{-i\theta + i\gamma}$, where $\gamma$ is an arbitrary constant. Clearly,  $\Psi_{vortex,\gamma} (\mathbf{r})$ and $\Psi_{vortex} (\mathbf{r})$ have the same energy. We may expect to have a Goldstone-like normal mode that shifts the phase while preserving the vortex position. This mode must necessarily occur in the $m=1$ sector as $\big(\Psi_{vortex,\gamma} (\mathbf{r}) - \Psi_{vortex} (\mathbf{r})\big)$ is proportional to $e^{-i\theta}$. However, we note that even an infinitesimal change in the phase corresponds to a long-ranged change in the order parameter. 
As the basis states (Landau levels) are localized wavefunctions, this physics can only be accessed if we retain all orders in the expansion.

In Appendix~\ref{app.spectrum}, we show that a gap persists in the $m=1$ sector as more and more Landau levels are included. The Goldstone-like mode presumably only appears when the Landau level cutoff is taken to infinity.

Our formalism based on a truncated expansion is not capable of addressing the idealized problem that should have two Goldstone-like modes.
In order to justify our approach, we invoke pinning mechanisms to preclude the two symmetries. We first discuss the disruption of translational symmetry. This can be achieved by a simple impurity potential. Indeed, pinning due to impurities is a well studied phenomenon in superconductors\cite{Matsushita2007,Woerdenweber2017}. where it is known to stabilize vortex lattices. In ultracold atomic gases, pinning can be naturally induced by the trapping potential. It can also be induced by an imposed potential, as studied in Ref.~\onlinecite{Reijnders2005}.
In such a setup, a smooth translation acquires an energy cost. In other words, the low-energy translational mode is neither long-ranged nor gapless. Rather, it acquires a length scale that is set by the pinning potential. In our calculations, we have not explicitly included a pinning potential. However, our choice of a cutoff in the Landau levels is equivalent to introducing a length scale in the problem: the higher the cutoff, the larger the length scale, leading to a broader low-energy mode. In other words, the truncation in the Landau level expansion encodes the strength of the pinning potential. We demonstrate this in Appendix~\ref{app.pinning}. 

We next consider phase pinning. This is easier to conceptualise in the case of a superconductor. A thin layer of a superconductor can be grown on a bulk superconductor with higher phase stiffness. An imposed magnetic field will create vortex lines that pierce both materials. The phase of the top layer will then be `pinned' to that of the substrate due to Josephson coupling. The physics of the layer can be probed selectively, e.g., using a beam of light to measure absorption. In this situation, the phase mode cannot be long-ranged and gapless. It acquires a length scale set by the strength of Josephson coupling. On the same lines, we argue that phase pinning can be achieved in a ultracold atomic gas. This can be realized in a two-component ultracold atomic gas where both components experience the same simulated magnetic field\cite{Kasamatsu2003}. Now, if one component is selectively probed, its phase will not show a gapless Goldstone-like mode.

While we do not explicitly include phase pinning in our calculations, we nonetheless have a length scale in the form of a cutoff in the Landau level expansion. This is reflected as a gap in the $m=1$ sector of the normal mode spectrum. Equivalently, it manifests as a finite spatial range of the low-energy phase mode. 

In summary, our results for the excitations about a single vortex are justified when two pinning mechanisms are present. The low energy excitations can then be viewed as analogues of atomic orbitals that are exponentially localized about the vortex centre. 
In Sec.~\ref{sec.tb} below, we use the single vortex spectra as building blocks to understand excitations in vortex lattices. The resulting tight binding dispersion is also justified only when pinning mechanisms are present.

\section{The vortex lattice and its periodicity}
\label{sec.VL}
To give a precise meaning to the notion of a vortex lattice, we define it as an equilibrium order parameter configuration where the \textit{amplitude} is periodic under discrete lattice translations. This is a physically meaningful definition as the amplitude is an observable quantity. The well known Abrikosov solution conforms to this definition, as its phase does not have spatial periodicity. 
We emphasize that the Abrikosov solution is a limiting case that is derived in the nearly-linear regime, where the parameter $b$ in Eq.~\ref{eq.psi_eq} is small. The discussion below holds for a general vortex lattice with no assumption of linearity. We now discuss the phase of the vortex lattice solution and its properties under translation.

We consider a two-dimensional lattice with primitive lattice vectors $\hat{a}$ and $\hat{b}$. These vectors are perpendicular to each other in a square vortex lattice, but not in the more common triangular lattice geometry. For the sake of simplicity, we assume that the two primitive lattice vectors are of the same length, $\ell$, which serves as the lattice constant. 
A generic lattice translation is denoted by $\mathcal{T}_{p,q}$, representing translation by $\mathbf{R}_{p,q} = p\hat{a} + q\hat{b}$, where $p$ and $q$ are integers. We take the order parameter to be given by $\Psi_{VL}(\mathbf{r})$, satisfying $\vert \Psi_{VL}(\mathbf{r}+\mathbf{R}_{p,q}) \vert = \vert \Psi_{VL}(\mathbf{r})\vert$.

As an equilibrium configuration, $\Psi_{VL}(\mathbf{r})$ is a solution of the Landau Ginzburg equation in Eq.~\ref{eq.psi_eq}. To see the effect of a lattice translation, we consider the same equation at a translated position, 
\bea
\nonumber &~& \Big[\frac{1}{2m^*}\big( -i\hbar\boldsymbol{\nabla} +2e\textbf{A} (\mathbf{r}-\mathbf{R}_{p,q})\big)^{2} + a \\
&+& 2b \vert \Psi_{VL}  (\mathbf{r}-\mathbf{R}_{p,q}) \vert ^{2} \Big]\Psi_{VL}  (\mathbf{r}-\mathbf{R}_{p,q}) = 0.~~~~~~
\label{eq.psi_eq_trans}
\eea 
As with Eq.~\ref{eq.psi_eq_shift} above, we now interpret this as a gauge transformation on Eq.~\ref{eq.psi_eq} using
\bea
\textbf{A} (\mathbf{r}-\mathbf{R}_{p,q}) = \textbf{A} (\mathbf{r}) + \boldsymbol{\nabla} \lambda_{p,q} (\mathbf{r}),
\label{eq.Atrans}
\eea
where $\lambda_{p,q} (\mathbf{r}) =  \frac{B}{2} \{\hat{z} \cdot \mathbf{r} \times \mathbf{R}_{p,q}\}$ is a scalar function. 
By undoing the gauge transformation in Eq.~\ref{eq.psi_eq_trans}, we find a new solution to Eq.~\ref{eq.psi_eq}, given by 
\bea
\Psi_{VL,p,q}(\mathbf{r}) \equiv e^{2ie \lambda_{p,q}(\mathbf{r})/\hbar}\Psi_{VL}  (\mathbf{r}-\mathbf{R}_{p,q}).
\eea
This new solution is obtained by first translating the reference vortex lattice solution and then attaching a gauge-derived phase. We note that $\Psi_{VL,p,q}(\mathbf{r})$ has the same order parameter amplitude as the reference solution. We argue that it does \textit{not} represent a distinct solution, but rather the same physical state with an additional constant global phase,
\bea
\Psi_{VL}  (\mathbf{r}) \equiv e^{-i\theta_{p,q}} e^{2ie \lambda_{p,q}(\mathbf{r})/\hbar}\Psi_{VL}  (\mathbf{r}-\mathbf{R}_{p,q}).~~
\label{eq.psi_trans_constraints}
\eea 
As seen from Eq.~\ref{eq.psi_eq}, a global phase can always be added to a given solution. This indicates that two solutions differing by a global phase represent the same physical solution. The global phase here, $\theta_{p,q}$, can depend on the translation vector, i.e., on $p$ and $q$. 

The action of lattice translation is captured by Eq.~\ref{eq.psi_trans_constraints} which gives the phase picked up by a vortex lattice under any lattice translation. As is well known, the set of all lattice translations forms a group. For example, a translation by $(p,q)$, following one by $(p',q')$, is equivalent to a net translation by $(p+p',q+q')$. This strongly constrains the phase angles $\theta_{p,q}$ (see details in Appendix~\ref{app.VL_trans}). It narrows down the allowed values of the phase to 
\bea
\theta_{p,q} = p \theta_a + q\theta_b + pq\pi,
\label{eq.theta_mn}
\eea
where $\theta_a$ and $\theta_b$ are phases associated with translations along $\hat{a}$ and $\hat{b}$ respectively.

To determine the phases $\theta_a$ and $\theta_b$, we first consider the linearized Landau Ginzburg equation, where the term proportional to $b$ in Eq.~\ref{eq.psi_eq} is ignored. In this regime, explicit expressions for vortex lattice solutions are known. The expression for a square vortex lattice was provided by Abrikosov\cite{Abrikosov1957}, while that for a triangular lattice was provided by Kleiner et. al.\cite{Kleiner1964} later. While these solutions correspond to the Landau gauge, they can be easily converted to the symmetric gauge to suit our discussion. We find that both the square and triangular solutions satisfy Eq.~\ref{eq.psi_trans_constraints}. We find that the additional phase, $\theta_{p,q}$, agrees with Eq.~\ref{eq.theta_mn} with $\theta_a = \theta_b = 0$. 
This completely determines the translational properties of the vortex lattice. 

We assert that the same translation-phase relations hold even in the presence of a non-linear term, due to the following topological argument. We postulate that the solution in the non-linear case can be adiabatically accessed from the linear regime, by smoothly increasing the parameter $b$ in Eq.~\ref{eq.psi_eq}. In the case of a square vortex lattice, we expect the order parameter to evolve smoothly from the Abrikosov solution, while preserving the lattice geometry. This constrains the angles $\theta_a$ and $\theta_b$ to remain fixed at zero during this evolution. Otherwise, there will be rapid long-ranged changes in the phase of the order parameter. For example, let us consider a lattice site corresponding to a large value of $p$ but with $q=0$. Upon increasing the strength of the non-linear term from zero, the $\theta_{p,q}$ component of the phase will rapidly change from $0$ (in the linear Abrikosov regime) to a large value, $p\theta_a$ (in the non-linear regime). This rapid change contradicts the postulate of adiabatic evolution. This problem does not arise if $\theta_a$ and $\theta_b$ always remain fixed at zero. We henceforth assume that the vortex lattice solution always satisfies Eq.~\ref{eq.psi_trans_constraints} with $\theta_a = \theta_b = 0$ in Eq.~\ref{eq.theta_mn}. The effect of lattice translations on the vortex lattice is then given by
\bea
\Psi_{VL}  (\mathbf{r}) \equiv e^{i pq\pi} e^{2ie \lambda_{p,q}(\mathbf{r})/\hbar}\Psi_{VL}  (\mathbf{r}-\mathbf{R}_{p,q}).~~
\label{eq.psi_trans_constraints_b}
\eea 
This form can be viewed as a representation of Zak's magnetic translation group\cite{Zak1,Zak2,Rosenstein2010}. 

\section{Excitations in the vortex lattice: Bloch's theorem}
\label{sec.bloch}
We now discuss collective excitations about a reference state that represents a vortex lattice. As the starting point, we consider a solution, given by $\{u(\mathbf{r}),v(\mathbf{r})\}$, to Eq.~\ref{eq.uveqs} with $\Psi_0(\mathbf{r}) = \Psi_{VL}(\mathbf{r})$. As the reference state forms a lattice, we expect the normal mode to be constrained by discrete translational symmetries. Below, we derive an analogue of Bloch's theorem that brings this out. To see the effect of a lattice translation, we consider Eq.~\ref{eq.uveqs} at a shifted position,
\begin{widetext}
\bea
\nonumber \Big[ \frac{1}{2m^*}\big( -i\hbar\boldsymbol{\nabla} + 2e\textbf{A}'(\mathbf{r})\big)^{2} + a + \epsilon + 2b\vert \Psi_{VL}(\mathbf{r}) \vert^{2} \Big]u(\mathbf{r}-\mathbf{R}_{p,q})
 - b\Psi_{VL}^{2}(\mathbf{r}-\mathbf{R}_{p,q}) v (\mathbf{r}-\mathbf{R}_{p,q}) =  0,\\
  \Big[ \frac{1}{2m^*}\big( -i\hbar\boldsymbol{\nabla} - 2e \textbf{A}'(\mathbf{r})\big)^{2} + a - \epsilon + 2b\vert\Psi_{VL}(\mathbf{r})\vert^{2} \Big]v(\mathbf{r}-\mathbf{R}_{p,q}) 
- b\Psi_{VL}^{*2}(\mathbf{r}-\mathbf{R}_{p,q}) u (\mathbf{r}-\mathbf{R}_{p,q})=  0,
\label{eq.Aprime}
\eea
where we have used the fact that the order parameter amplitude is invariant under a lattice translation. 
We denote $\textbf{A}'(\mathbf{r}) = \textbf{A}(\mathbf{r}-\mathbf{R}_{p,q}) = \textbf{A}(\mathbf{r})+\boldsymbol{\nabla} \lambda_{p,q}(\mathbf{r})$, with $\lambda_{p,q}(\mathbf{r})$ defined above in Eq.~\ref{eq.Atrans}. The change in the vector potential suggests that these equations can be viewed as a gauge transformation. We proceed to undo this gauge transformation to recover a different solution to Eq.~\ref{eq.uveqs}. At the same time, we substitute for $\Psi_{VL}(\mathbf{r}-\mathbf{R}_{p,q})$ using Eq.~\ref{eq.psi_trans_constraints_b}. We obtain
\bea
\nonumber \Big[ \frac{1}{2m^*}\big( -i\hbar\boldsymbol{\nabla} + 2e\textbf{A}(\mathbf{r})\big)^{2} + a + \epsilon + 2b\vert \Psi_{VL}(\mathbf{r}) \vert^{2} \Big] e^{2ie\lambda_{p,q}(\mathbf{r})/\hbar}
u(\mathbf{r}-\mathbf{R}_{p,q}) 
- b\Psi_{VL}^{2}(\mathbf{r})
e^{-2ie\lambda_{p,q}(\mathbf{r})/\hbar} v (\mathbf{r}-\mathbf{R}_{p,q}) &=&  0\\
\Big[ \frac{1}{2m^*}\big( -i\hbar\boldsymbol{\nabla} - 2e\textbf{A}(\mathbf{r})\big)^{2} + a - \epsilon + 2b\vert\Psi_{VL}(\mathbf{r})\vert^{2} \Big] e^{-2ie\lambda_{p,q}(\mathbf{r})/\hbar}  
v(\mathbf{r}-\mathbf{R}_{p,q}) - b\Psi_{VL}^{*2} (\mathbf{r}) e^{2ie\lambda_{p,q}(\mathbf{r})/\hbar} 
u (\mathbf{r}-\mathbf{R}_{p,q})&=& 0.~~~~~~
\label{eq.shifted_uv}
\eea
\end{widetext}
We have now arrived at a different solution to Eq.~\ref{eq.uveqs} given by $\{  e^{2ie\lambda_{p,q}(\mathbf{r})/\hbar} u(\mathbf{r}-\mathbf{R}_{p,q}),e^{-2ie\lambda_{p,q}(\mathbf{r})/\hbar} v(\mathbf{r}-\mathbf{R}_{p,q})\}$. Clearly, this solution is related to $\{u(\mathbf{r}),v(\mathbf{r})\}$, the original solution, by a lattice translation followed by phase attachment. We argue that the two represent the same physical normal mode. This implies that they can differ by utmost a global phase. To see this, we note that Eq.~\ref{eq.uveqs} allows for an arbitrary global phase to be added to any given solution, $ \{u(\mathbf{r}),v(\mathbf{r})\} \rightarrow  \{e^{i\phi}u(\mathbf{r}),e^{i\phi}v(\mathbf{r})\}$. As a consequence, two modes that differ by a global phase represent the same physical solution. Here, we take the global phase to be $\phi=\beta_{p,q}$ as it can depend on $\mathbf{R}_{p,q}$, the lattice translation vector. We have
\bea
\nonumber
u(\mathbf{r})  &=&  e^{-i\beta_{p,q}} e^{2ie\lambda_{p,q}(\mathbf{r})/\hbar} u (\mathbf{r}-\mathbf{R}_{p,q}), \\
v(\mathbf{r}) &=& e^{-i\beta_{p,q}}  e^{-2ie\lambda_{p,q}(\mathbf{r})/\hbar} v (\mathbf{r}-\mathbf{R}_{p,q}).
\label{eq.uvtrans_phase}
\eea
These relations capture the transformation of the normal modes under lattice translations. It remains to determine $\beta_{p,q}$, the phase accrued by the normal mode. We do this by using the closure properties of the lattice translation group whereby $\mathcal{T}_{p,q}\mathcal{T}_{p',q'}\equiv \mathcal{T}_{p+p',q+q'}$. This leads to the same structure as discussed in the context of the vortex lattice solution in Eq.~\ref{eq.psi_trans_constraints}. Relegating the details to Appendix~\ref{app.mode_trans}, we find
\bea
\beta_{p,q} = \mathbf{k}\cdot\mathbf{R}_{p,q} + pq\pi,
\label{eq.beta_mn}
\eea
where $\mathbf{k}$ can be identified as a momentum vector in the first Brillouin zone of the (vortex) lattice.  

From the form of $\beta_{p,q}$, we see that each normal mode corresponds to a `crystal momentum', $\mathbf{k}$. Modes with different $\mathbf{k}$'s are independent. We will see this explicitly below, in a tight binding context.
These arguments demonstrate that $\mathbf{k}$ serves as a good quantum number for normal modes. To reflect this, we introduce a momentum subscript for ${u}(\mathbf{r})$ and $v(\mathbf{r})$. We arrive at the appropriate form of Bloch's theorem for normal modes in a vortex lattice,
\bea
\nonumber u_{\mathbf{k}} (\mathbf{r}-\mathbf{R}_{p,q}) &=& e^{-2ie \lambda_{p,q}(\mathbf{r})/\hbar}e^{i\mathbf{k} \cdot \mathbf{R}_{p,q} + i pq\pi} u_{\mathbf{k}} (\mathbf{r}), \\
v_{\mathbf{k}} (\mathbf{r}-\mathbf{R}_{p,q}) &=& e^{2ie \lambda_{p,q}(\mathbf{r})/\hbar}e^{i\mathbf{k} \cdot \mathbf{R}_{p,q} + i pq\pi} v_{\mathbf{k}} (\mathbf{r}).
\label{eq.Bloch}
\eea
This completely determines the action of a lattice transformation on a normal mode. As with the traditional statement of Bloch's theorem, it is helpful to recast this as an explicit form for a given normal mode\cite{AshcroftMermin}. 
We present an ansatz,
\bea
\nonumber {u}_{\mathbf{k}} (\mathbf{r}) \equiv \sum_{p,q} \phi_\mathbf{k} (\mathbf{r}-\mathbf{R}_{p,q})
 e^{-i \mathbf{k}\cdot \mathbf{R}_{p,q}} e^{2ie \lambda_{p,q}(\mathbf{r})/\hbar} e^{i pq\pi},~~\\
 {v}_{\mathbf{k}} (\mathbf{r}) \equiv \sum_{p,q} \xi_\mathbf{k} (\mathbf{r}-\mathbf{R}_{p,q})
 e^{-i \mathbf{k}\cdot \mathbf{R}_{p,q}} e^{-2ie \lambda_{p,q}(\mathbf{r})/\hbar} e^{i pq\pi}.~~
\label{eq.ansatz}
\eea
Here, the functions $\phi_\mathbf{k}(\mathbf{r})$ and $\xi_\mathbf{k}(\mathbf{r})$ are the analogues of Wannier functions. We will interpret them as combinations of atomic orbitals in the following section. 
Crucially, the ansatz in Eq.~\ref{eq.ansatz} satisfies Bloch's theorem as given in Eq.~\ref{eq.Bloch}, as can be checked by direct substitution. It is also particularly suited to a tight binding approach, as we show below. 

\section{Tight binding approach}
\label{sec.tb}
Having arrived at a suitable form of Bloch's theorem, we seek to develop a tight binding approach. 

\subsection{Solutions near lattice sites as atomic orbitals }
We assume that the origin is one of the zeroes of the vortex lattice solution, i.e., the origin is one of the sites of the vortex lattice. In the vicinity of the origin, at distances smaller than the lattice spacing (denoted by $\ell$), we assume that the vortex lattice solution resembles that of an isolated vortex,
\bea
\Psi_{VL} (\mathbf{r}) \approx \Psi_{vortex} (\mathbf{r}), ~~\mathbf{r}\ll \ell,
\eea  
Fluctuations within this region can be thought of as those around an isolated vortex. 
We express such a solution as $(u_0(\mathbf{r}),v_0(\mathbf{r}))$, assuming that the fluctuations are as given by Eq.~\ref{eq.etaform}.
We note that both $u_0(\mathbf{r})$ and $v_0(\mathbf{r})$ are centred at the origin.
As discussed above, these functions can be written in the Landau level basis.

Likewise, in the vicinity of a different lattice site, the vortex lattice resembles the profile of an isolated vortex. However, there is a nuance here. We consider a point in the vicinity of a site $\mathbf{R}_{p,q}$ given by $\mathbf{r} = \mathbf{R}_{p,q} + \delta\mathbf{r}$, where $\delta\mathbf{r}\ll \ell$. Using Eq.~\ref{eq.psi_trans_constraints_b}, we have
\bea
\nonumber \Psi_{VL}  (\mathbf{r}) &=& \nonumber \Psi_{VL}  (\mathbf{R}_{p,q} + \delta\mathbf{r}) = \Psi_{VL}  (\delta\mathbf{r}) e^{-i pq\pi}  e^{2ie \lambda_{p,q}(\delta\mathbf{r})/\hbar} \\ 
\nonumber &\approx&  \Psi_{vortex} (\delta\mathbf{r}) e^{-i pq\pi}  e^{2ie \lambda_{p,q}(\delta\mathbf{r})/\hbar},\delta\mathbf{r} \ll \ell ~~~\\
\nonumber &=&  \Psi_{vortex} (\mathbf{r} - \mathbf{R}_{p,q})e^{-i pq\pi}  e^{2ie \lambda_{p,q}(\mathbf{r})/\hbar}, \\
&\equiv&  \Psi_{vortex,p,q} (\mathbf{r}) .
\label{eq.psi_neighbourhood}
\eea
We have used $\lambda_{p,q}(\delta\mathbf{r}) = \lambda_{p,q}(\mathbf{r})$, a relation that follows from the definition of $\lambda$ given below Eq.~\ref{eq.Atrans}. In the last step, we have defined $ \Psi_{vortex,p,q} (\mathbf{r})$ as the approximate functional form of $\Psi_{VL}$ in the vicinity of $\mathbf{R}_{p,q}$. In amplitude, it represents a shifted isolated vortex, centred at $\mathbf{R}_{p,q}$. 
However, it carries an additional ($p,q$)-dependent phase that originates from the gauge structure. This phase is inherited by the fluctuation modes in this vicinity. By suitably adapting the arguments used in Eqs.~\ref{eq.Aprime} and \ref{eq.shifted_uv} above, we see that a fluctuation in this neighbourhood is given by 
\bea
\nonumber  (u_0(\mathbf{r}-\mathbf{R}_{p,q}) e^{2ie \lambda_{p,q}(\mathbf{r})/\hbar}    &,&
 v_0(\mathbf{r}-\mathbf{R}_{p,q}) e^{-2ie \lambda_{p,q}(\mathbf{r})/\hbar} )\\
&\times& e^{i pq\pi} .
\label{eq.uv_mn}
\eea 
Here, we have added an overall phase of $(pq\pi)$ for later convenience. As argued below Eq.~\ref{eq.shifted_uv}, we are free to add a constant overall phase to a given solution. In Eq.~\ref{eq.uv_mn}, we have arrived at a fluctuation that is centred at $\mathbf{R}_{p,q}$. 
We see that the $u_0$ and $v_0$ functions pick up opposite site-dependent phases, apart from an overall phase. We recognize these as the same phases that appear in Eq.~\ref{eq.ansatz}, the ansatz based on Bloch's theorem. We now interpret $u_0(\mathbf{r})$ and $v_0(\mathbf{r})$ as the Wannier functions, $\phi_{\mathbf{k}}$ and $\xi_{\mathbf{k}}$, respectively. Given an arbitrary site, the corresponding Wannier functions can be obtained by directly translating the Wannier functions centred at the origin, without attaching any additional phases.

\subsection{Linear combination of atomic orbitals approach}

The normal modes in a vortex lattice have been shown to obey an analogue of Bloch's theorem. A suitable form for the normal mode is given in Eq.~\ref{eq.ansatz} in terms of Wannier functions, $\phi_{\mathbf{k}}$ and $\xi_{\mathbf{k}}$. We now represent the Wannier functions in terms of atomic orbitals, i.e., normal modes around an isolated vortex. We first consider a vortex centred at the origin. Following Sec.~\ref{sec.isolated}, its low energy atomic orbitals can be well approximated by
\bea
u (\mathbf{r}) = u_{m} (\zeta_{0,m}^{LL} (\mathbf{r}))^*; ~~v (\mathbf{r}) = v_{2-m} \zeta_{0,2-m}^{LL} (\mathbf{r}).
\eea
Taking the origin to be a lattice site, the corresponding Wannier functions can be taken to be composed of all such atomic orbitals, i.e., we take $\phi_\mathbf{k}$ and $\xi_\mathbf{k}$ to be linear combinations of the local $u$'s and $v$'s respectively. Depending on the accuracy required, any number of atomic orbitals can be included. In the interest of simplicity, we restrict ourselves to the four lowest-energy atomic orbitals. As described in Sec.~\ref{sec.isolated} above, these correspond to two gyrotopic modes with $m=0$ and two breathing modes with $m=1$.
We write
\bea
\left(
\begin{array}{c}
\phi_{\mathbf{k}} (\mathbf{r}) \\
\xi_{\mathbf{k}} (\mathbf{r})
\end{array}
\right) 
\equiv
\sum_{j=1}^4  b_j \eta_j^{0,0} (\mathbf{r})
= 
\sum_{j=1}^4  b_j  \left(
\begin{array}{c}
u_j (\mathbf{r}) \\
 v_j (\mathbf{r})
\end{array}
\right). 
\label{eq.Wannierzero}
\eea
where $j$ sums over the four lowest modes: $j=1,2$ correspond to gyrotropic modes with $m=0$ while $j=3,4$ correspond to breathing modes with $m=1$. We have used $\eta_j^{0,0}(\mathbf{r})$ to denote each two-component atomic orbital. The $(0,0)$ superscript in $\eta$ indicates that these orbitals are centred at the origin.
Explicit expressions for $u_j (\mathbf{r})$ and $v_j (\mathbf{r})$ are given in Appendix~\ref{app.spectrum}. The coefficients $b_j$ determine the precise combinations of atomic orbitals that serve as Wannier functions. They will be determined below as solutions of an eigenvalue problem.

In Eq.~\ref{eq.Wannierzero}, we have expressed the Wannier function at the origin in terms of atomic orbitals. The Wannier functions at other sites can be obtained by a direct translation (without any additional phases) by the corresponding lattice translation vector.
 
\subsection{Tight binding construction}
The fluctuation normal modes about a vortex lattice are to be found from Eq.~\ref{eq.uveqs}, taking $\Psi_0 (\mathbf{r}) \equiv \Psi_{VL}(\mathbf{r})$. This represents an eigenvalue equation with a `Hamiltonian' matrix. We consider this equation in the vicinity of $\mathbf{R}_{p,q}$, a given lattice site. Using Eq.~\ref{eq.psi_neighbourhood}, we express the vortex lattice solution here as
\bea
\nonumber \Psi_{VL}(\mathbf{r})  = \Psi_{vortex,p,q}(\mathbf{r}) + \delta \Psi_{p,q} (\mathbf{r}).
\eea 
 The first term represents the profile of an isolated vortex centred at $\mathbf{R}_{p,q}$, but with additional gauge-derived phases (see Eq.~\ref{eq.psi_neighbourhood}). The second represents the deviation from this approximate form. The deviation is small in the near vicinity of $\mathbf{R}_{p,q}$, becoming appreciable over distances of the order of the lattice spacing.

We now divide the Hamiltonian matrix of Eq.~\ref{eq.uveqs} as 
\bea
H \approx H_{p,q} + \delta H_{p,q},
\label{eq.Hsplit}
\eea
where 
\begin{widetext}
\bea
H_{p,q} = \left(
\begin{array}{cc}
-\frac{1}{2m^*}\big( -i\hbar\boldsymbol{\nabla} +2e \textbf{A}(\mathbf{r})\big)^{2} - a - 2b\vert \Psi_{vortex,p,q} (\mathbf{r}) \vert^{2} &  b\Psi_{vortex,p,q}^{2} (\mathbf{r}) \\
 -b\Psi_{vortex,p,q}^{*2}(\mathbf{r}) & \frac{1}{2m^*}\big(- i\hbar\boldsymbol{\nabla} - 2e\textbf{A}(\mathbf{r})\big)^{2} + a + 2b\vert \Psi_{vortex,p,q} (\mathbf{r}) \vert^{2}
\end{array}
\right)
\eea
is the Hamiltonian corresponding to an isolated vortex at $\mathbf{R}_{p,q}$, representing the `atomic' Hamiltonian. The deviation is given by 
\bea
\delta H_{p,q} = \left(
\begin{array}{cc}
-2b (\vert \Psi_{VL} (\mathbf{r}) \vert^2 -  \vert \Psi_{vortex,p,q} (\mathbf{r}) \vert^2  ) &  b( \Psi_{VL}^{2} (\mathbf{r}) - \Psi_{vortex,p,q}^{2} (\mathbf{r}) )\\
 -b(\Psi_{VL}^{*2}(\mathbf{r})-\Psi_{vortex,p,q}^{*2}(\mathbf{r})) & 2b   (\vert \Psi_{VL} (\mathbf{r}) \vert^2 -  \vert \Psi_{vortex,p,q} (\mathbf{r}) \vert^2  )
\end{array}
\right).
\label{eq.deltaH}
\eea
\end{widetext}
The eigenvectors of $H_{p,q}$ can be written using Eq.~\ref{eq.uv_mn}, using results from the isolated vortex problem in Sec.~\ref{sec.isolated}. We have
\bea
H_{p,q} \eta_{j}^{p,q} (\mathbf{r})  =\epsilon_j \eta_{j}^{p,q} (\mathbf{r}),
\eea
where $\eta_j^{p,q}(\mathbf{r})$ and $\epsilon_j$ represent an eigenvector and the corresponding eigenvalue of the isolated vortex problem. In explicit form, the eigenvector is given by 
\bea
\eta_{j}^{p,q} (\mathbf{r}) = \left(
\begin{array}{c}
u_j (\mathbf{r} - \mathbf{R}_{p,q}) e^{2ie \lambda_{p,q}(\mathbf{r})/\hbar}  e^{ipq\pi} \\
v_j (\mathbf{r} - \mathbf{R}_{p,q}) e^{-2ie \lambda_{p,q}(\mathbf{r})/\hbar} e^{ipq\pi}
\end{array}
\right),
\label{eq.eta}
\eea
where $u_j$ and $v_j$ represent the solution for an isolated vortex at the origin (see Appendix~\ref{app.spectrum} for explicit expressions).
Combining Eqs.~\ref{eq.ansatz}, \ref{eq.Wannierzero} and \ref{eq.eta}, the Bloch normal mode for the vortex lattice can be written as
\bea
\left(
\begin{array}{c}
u_\mathbf{k} (\mathbf{r}) \\
v_\mathbf{k} (\mathbf{r}) 
\end{array}
\right)  = \sum_{p,q} \sum_{j=1}^4 b_j e^{-i\mathbf{k}\cdot\mathbf{R}_{p,q}} \eta_j^{p,q} (\mathbf{r}).
\eea
We take this form to be an eigenvector of $H$ with eigenvalue $\varepsilon_\mathbf{k}$. We express the Hamiltonian as in Eq.~\ref{eq.Hsplit}, splitting it into two parts: the `atomic Hamiltonian' at site $(p',q')$ and the deviation from it. We obtain
\bea
\big(H_{p',q'} + \delta H_{p',q'} - \varepsilon_\mathbf{k}\big) \sum_{p,q} \sum_{j=1}^4 b_j e^{-i\mathbf{k}\cdot\mathbf{R}_{p,q}} \eta_j^{p,q} (\mathbf{r}) =0.~~~
\eea
To extract a tight-binding form, we left-multiply this equation by $\big\{\big(\eta_{j'}^{p',q'}(\mathbf{r})\big)^\dagger \sigma_z\big\}$ and integrate over all space. The Pauli matrix, $\sigma_z$, is necessary to invoke orthonormality of atomic orbitals following the orthonormality relation given below Eq.~\ref{eq.uveqs}.  
We obtain a set of four equations, one for each value of $j'$, given by
\begin{widetext}

\bea
\nonumber b_{j'} (-\mathrm{sign}(\epsilon_{j'})\epsilon_{j'}-\varepsilon_\mathbf{k}) &+&
 \sum_{j} b_j {s}_{j',j} (p',q') \\
&+& \sum_{(p,q) \neq (p',q')} 
e^{-i\mathbf{k}\cdot\mathbf{R}_{p-p',q-q'} }
\sum_{j=1}^4  
  b_j 
  \Big\{ 
(\epsilon_{j'}-\varepsilon_\mathbf{k})  t_{j',j}(p',q',p,q)
  + r_{j',j}(p',q',p,q) ) 
 \big\} =  0.~~~
 \label{eq.bjmatrix}
\eea
We have defined three types of overlap integrals,
\bea
s_{j',j} (p',q') &\equiv&  \int d^2 r \big(\eta_{j'}^{p',q'}(\mathbf{r})\big)^\dagger \sigma_z (\delta H_{p',q'}) \eta_{j}^{p',q'}(\mathbf{r}),  \label{eq.s_def} \\
t_{j',j}(p',q',p,q) &\equiv&  \int d^2 r \big(\eta_{j'}^{p',q'}(\mathbf{r})\big)^\dagger \sigma_z \eta_{j}^{p,q}(\mathbf{r}),  \label{eq.t_def} \\
 r_{j',j}(p',q',p,q) &\equiv&  \int d^2 r \big(\eta_{j'}^{p',q'}(\mathbf{r})\big)^\dagger \sigma_z (\delta H_{p',q'}) \eta_{j}^{p,q}(\mathbf{r}).
 \label{eq.r_def}
\eea 
\end{widetext}
These are direct analogues of the integrals that appear in the traditional tight binding scheme for electronic bands\cite{AshcroftMermin}. The first is a single-site term, $s_{j',j}(p',q') $. It represents the matrix element of the deviation term ($\delta H_{p',q'}$) between atomic orbitals. The second term represents an overlap integral between atomic orbitals on different sites, e.g., on nearest neighbour sites. The third represents the matrix element of the deviation term between atomic orbitals centred on different sites.

This problem embodies a crucial difference with respect to the standard formulation of tight binding for electronic bands. Here, the atomic orbitals differ from one site to the next due to phases that arise from the gauge structure. In order to have a consistent tight binding formulation, the overlap integrals must satisfy the following properties: (i) The on-site overlap $s_{j',j}(p',q')$ must take the same value on every site. If this holds, we may drop the $p'$ and $q'$ arguments and denote the integral as $s_{j',j}$. (ii) We have two types of inter-site overlap integrals, $t_{j',j}(p',q',p,q)$ and $r_{j',j}(p',q',p,q)$. They are defined in terms of atomic orbitals on two sites, $(p,q)$ and $(p',q')$. Their value must depend solely on the separation between the sites, i.e., on $(p-p', q-q')$ and not on the sites themselves. If this is true, we may write $t_{j',j}(p',q',p,q) \equiv t_{j',j}(p-p',q-q')$ and $r_{j',j}(p',q',p,q) \equiv r_{j',j}(p-p',q-q')$.
Indeed, these two properties hold true for the overlap integrals. We show this explicitly in Appendix~\ref{app.overlap_trans}.

With the two properties mentioned above, Eq.~\ref{eq.bjmatrix} becomes independent of $(p',q')$. It reduces to a smaller set of independent equations, with one for each value of $j$. We view this is as a homogeneous system of linear equations in the $b_j$'s. We express this as a matrix equation
\bea
\sum_i M_{ij} (\mathbf{k},\varepsilon_\mathbf{k}) b_j = 0.
\eea
The matrix $M_{ij}$ depends on the momentum $\mathbf{k}$ as well as the as-yet-undetermined energy, $\varepsilon_\mathbf{k}$. With some careful arguments, it can be shown that this  matrix is Hermitian with real eigenvalues. In order to have a non-trivial solution, the determinant of $M_{ij} (\mathbf{k},\varepsilon_\mathbf{k}) $ must vanish. This leads to a polynomial equation in $\varepsilon_\mathbf{k}$ whose roots represent the band energies. At each root, one of the eigenvalues of $M_{ij} (\mathbf{k},\varepsilon_\mathbf{k})$ vanishes, forcing the determinant to vanish. The corresponding eigenvector gives the values of the $b_j$ coefficients, revealing the composition of the normal mode in terms of atomic orbitals.

We have discussed the general formulation of the tight binding scheme for normal modes in a vortex lattice. Below, we will take up a simple example and work out the band structure. At this stage, we list the required pieces of information that go into the tight binding calculation: 
\begin{enumerate}
\item  The strength of pinning mechanisms. This enters at the level of an isolated vortex, where it is reflected in the choice of cutoff in the Landau level expansion.

\item The profile of an isolated vortex as described by the ratio $\xi/\ell_B$: This quantity enters in the calculation of the normal mode spectrum about an isolated vortex. It determines atomic orbital eigenvalues and eigenvectors which are key ingredients in the tight binding scheme.
\item The geometry of the vortex lattice: This strongly constrains the overlap integrals. For example, we consider the four nearest neighbour bonds on the square lattice. If the $t$ and $r$ overlap integrals are known on one of the bonds, this immediately determines their values on the other three. This can be seen in Appendix~\ref{app.tb_rotations}, where we provide the values of the overlaps on the square lattice. 
We note that the lattice constant is fixed in terms of the magnetic length, $\ell_B$. This follows from the requirement that each unit cell (each vortex) must enclose one flux quantum.

\item The precise form of the vortex lattice solution: This is analogous to the periodic nuclear potential seen by an electron in the traditional tight binding description of electronic bands. It enters in the Hamiltonian deviation matrix ($\delta H$). Thereby, it determines the $s$ and $r$ overlap integrals defined above. In principle, the vortex lattice solution can be determined from experimental measurements or by large scale numerical minimization of the Landau Ginzburg free energy.
\end{enumerate}
We also note that the tight binding construction can be drastically simplified by limiting the inter-site overlaps to a few nearest neighbours. As shown in Sec.~\ref{sec.isolated}, the atomic orbitals are exponentially localized at their respective sites. As a consequence, the $t$ and $r$ overlaps decrease systematically with increasing bond length. A reasonable tight binding description may be obtained by restricting to nearest neighbours alone.

\subsection{Band structure in the square vortex lattice}
We now take the case of a square vortex lattice as an example. We assume an ersatz form for the vortex lattice solution that will enable us to carry out the tight binding analysis. We take its amplitude to be given by 
\bea
\nonumber \vert \Psi_{VL,sq} (\mathbf{r}) \vert^2 &\equiv& \frac{4\Delta_0^2}{\pi}\bigg\{ \sin^2\big(\sqrt{\frac{\pi}{2}}x\big) + \sin^2\big(\sqrt{\frac{\pi}{2}}y\big) \\
\nonumber &-& 0.5\sin^4 \big(\sqrt{\frac{\pi}{2}}x\big) -0.5\sin^4 \big(\sqrt{\frac{\pi}{2}}y\big) \\ 
&-& 0.35\sin^2 \big(\sqrt{\frac{\pi}{2}}x \big)\sin^2 \big(\sqrt{\frac{\pi}{2}}y\big)\bigg\}.
\label{eq.sqVLform}
\eea
Here, $\Delta_0 = \sqrt{-a/b}$ is the order parameter amplitude in the uniform superconductor. 
This ansatz has zeros on a square array, given by $\sqrt{2\pi}(p,q)$, where $p$ and $q$ are integers. Here, we have taken the unit of distance to be $\ell_B$ (magnetic length), so that the lattice constant comes out to be $\sqrt{2\pi}\ell_B$. This is consistent with requirement of a unit flux quantum passing through each plaquette. The same lattice constant appears in Abrikosov's exact form for a square vortex lattice. In the vicinity of each zero, this form encodes a linear increase in the superconducting amplitude. We have chosen the coefficients of the $\sin^2(.)$ terms so that the slope matches that of an isolated vortex with correlation length, $\xi = \frac{\ell_{B}}{2}$ -- an arbitrary value chosen for concreteness. 
We have chosen the coefficients of the quartic $\sin^4(.)$ terms so that the maximum value of $\vert \Psi_{VL,sq} (\mathbf{r}) \vert$ is less than $\Delta_0$ everywhere.

The phase of the vortex lattice solution is strongly constrained by the following considerations. The variation in phase from one unit cell to the next is fixed by Eq.~\ref{eq.psi_trans_constraints_b}. In addition, each lattice site represents a bonafide vortex. This imposes a winding of $2\pi$ in phase upon traversing a loop that encloses a single site. In order to evaluate the tight binding overlaps, we make reasonable assumptions that are consistent with these considerations. In Appendix~\ref{app.tb_rotations}, we describe the evaluation of overlap integrals and tabulate their values. For simplicity, we restrict the $t$ and $r$ overlaps to nearest neighbour bonds, setting all further overlaps to zero. We obtain the resulting $4\times 4$ $M_{ij}(\mathbf{k},\varepsilon_\mathbf{k})$ matrix (see Appendix~\ref{app.tb_rotations} for its explicit form). At each $\mathbf{k}$ of the square lattice Brillouin zone, we solve for $\varepsilon_\mathbf{k}$ by setting $\mathrm{Det}(M_{ij}(\mathbf{k},\varepsilon_\mathbf{k}))=0$.

\begin{figure}
\includegraphics[width=\columnwidth]{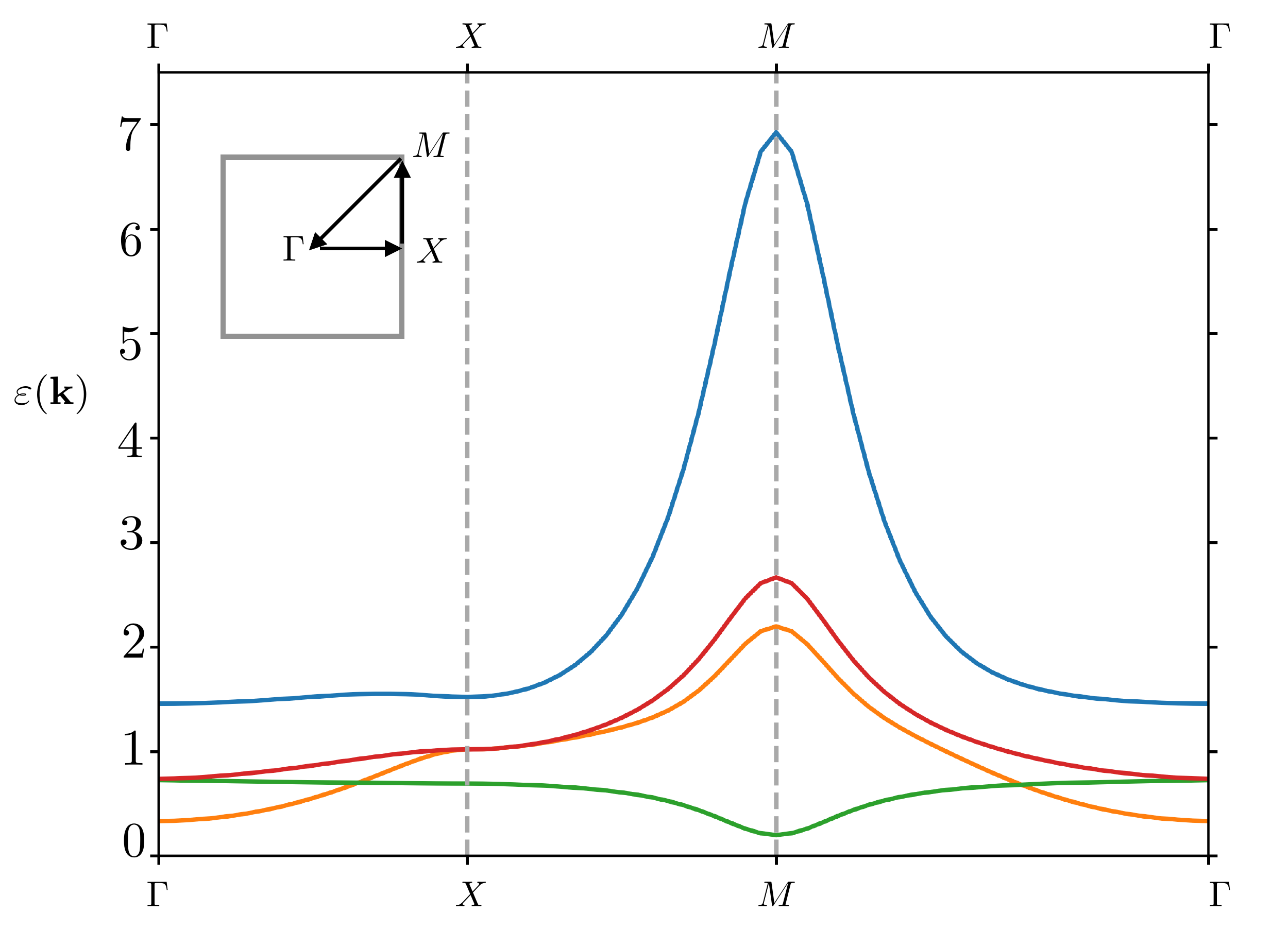}
\caption{Band structure of collective modes about a square vortex lattice. The energy is given in units of the Landau-Ginzburg parameter $a$.}
\label{fig.bs_square}
\end{figure}

The resulting band structure for the normal modes is shown in Fig.~\ref{fig.bs_square}. The energies are shown in units of $a$, the (Landau Ginzburg) parameter that has dimensions of energy. This quantity naturally emerges as the scale in the tight binding matrix. 
We note that all band energies are positive, reflecting stability of the vortex lattice. We have four bands as we have only retained four atomic orbitals. The `atomic limit' can be accessed by tuning all overlap integrals ($s$, $t$ and $r$) to zero. In this limit, we recover four flat bands, with the same energies as that obtained from the isolated vortex calculation. This allows us to identify each band as having breathing or gyrotropic character. As we approach the physical limit by reinstating the overlaps, we see interesting band crossing phenomena. For instance, in Fig.~\ref{fig.bs_square}, the lowest energy excitation occurs at the $M$ point ($\pi,\pi$ in the Brillouin zone). This excitation is purely gyrotropic in character. However, the same band (green line in the figure) has purely breathing mode character at the $\Gamma$ point. We see similar crossovers in all bands. For example, the band with the highest energy (blue curve in the figure) is gyrotropic in character at the Brillouin zone centre. However, at the M point, this mode is composed of breathing modes. This reflects mixing phenomena associated with band crossings.

\section{Consequences for experiments and numerical studies}
We have presented a study of collective modes in vortices and vortex lattices. Our results can be tested in several ways as we discuss below. 

\subsection{Collective modes about an isolated vortex}
\label{ssec.expt_isolated}
We first focus on our results for normal modes about an isolated vortex. We have shown that collective modes emerge with `m' as a quantum number, which can be viewed as angular momentum. 
We have derived the forms of these low energy modes using (a) an approximate ($\tanh$) form of the reference wavefunction and (b) a truncated Landau level expansion, where the cutoff is assumed to mimic the effect of a pinning potential. We assert that existence of distinct normal modes does not depend strongly on these assumptions. For example, if an explicit pinning potential is included, it will still lead to localized solutions with low energies. This is a generic property of low energy excitations in a medium with a point-like defect.

We have focussed on gyrotropic ($m=0$) and breathing ($m=1$) modes as low-energy excitations. In addition, we have other excitations with higher angular momentum ($m>2$). We propose simulated light absorption as a suitable approach to detect these modes experimentally. As a prerequisite, we assume that suitable pinning mechanisms are in place. For example, in order to have localized gyrotropic modes, we require a potential to pin the position of the vortex. In order to have localized breathing modes, we require pinning of the phase. 
To detect these modes, we propose using light with suitable choices of polarization and orbital angular momentum. In particular, we show that gyrotropic modes can be excited by using circularly polarized light. Breathing modes can be excited with `twisted' light with unit orbital angular momentum. In general, light with a certain orbital angular momentum excites normal modes that carry the same angular momentum.

We consider simulating a light beam impinging on the sample. The beam is taken to have a small amplitude so that it can be treated as a weak perturbation. It is encoded in a vector potential $\textbf{A}_{beam}(\textbf{r}, t)$ that is `seen' by the superfluid. We further assume that this light is turned on at time $t=0$ when the order parameter is given by $\psi_{vortex} (\mathbf{r})$, the equilibrium isolated-vortex solution. At $t>0$, the time-dependent Landau Ginzburg equation of Eq.~\ref{eq.SE} takes the form
\begin{eqnarray}
\nonumber i\hbar \frac{\partial\psi}{\partial t} &\approx & \frac{1}{2m^{*}}[ -i\hbar\boldsymbol{\nabla} + 2e\textbf{A}(\mathbf{r})]^{2}\psi  + a\psi + b|\psi|^{2}\psi \\ 
&+& \frac{2e}{m^{*}}\textbf{A}_{beam}(\mathbf{r},t).\Big[ - i\hbar\boldsymbol{\nabla} + 2e\textbf{A}(\mathbf{r}) \Big]\psi.
\label{eq.psi_pert}
\end{eqnarray}
Here, $\textbf{A}(\mathbf{r})$ is the reference vector potential corresponding to a uniform magnetic field and $\textbf{A}_{beam}(\mathbf{r},t)$ is the perturbation. 
We have negected $\mathcal{O}({\textbf{A}_{beam}}^2)$ terms as the amplitude of light is small. We consider the change in the order parameter after an infinitesimal time $\delta t$. From Eq.~\ref{eq.psi_pert}, we find 
\begin{eqnarray}
\nonumber \delta \psi (\mathbf{r},\delta t) &=& i(\delta t)\frac{2e }{m^{*}}\textbf{A}_{beam} (\mathbf{r},t)\cdot\Big[ - i\hbar\boldsymbol{\nabla} + 2e\textbf{A}(\mathbf{r}) \Big]\times \\
\nonumber &~&\psi_{vortex} (\mathbf{r}) \\
&=& i(\delta t) \textbf{A}_{beam} (\mathbf{r},t) \cdot \{ J (r) ~\hat{\theta} + i~ I (r) ~\hat{r} \} e^{-i \theta}.
\label{eq.deltapsi}
\end{eqnarray}
In the second line, we have acted the gauge-covariant current operator given by $\frac{2e}{m^{*}}\Big[ - i\hbar\boldsymbol{\nabla} + 2e\textbf{A} \Big]$ on the reference solution, $\psi_{vortex}(\mathbf{r})$. The resulting form is constrained by the form of the isolated vortex solution, $\Delta_{vortex}(\mathbf{r}) = \Delta (r) e^{-i\theta}$. It gives 
$\{ J (r) ~\hat{\theta} + i~ I (r) ~\hat{r} \} e^{-i \theta}$, where $J(r)$ and $I(r)$ are purely real functions that depend solely on the radial coordinate.

To determine the change in the order parameter, we consider the explicit form of $\textbf{A}_{beam}(\mathbf{r},t)$ given by\cite{Quinteiro2015}
\begin{eqnarray}
\nonumber \textbf{A}_{beam} (\mathbf{r},t) &=& F(r) \cos[\omega t  - (\kappa+\sigma) \theta ]~\hat{r} \\
&+&
\sigma F(r) \sin[\omega t  - (\kappa+\sigma) \theta] ~\hat{\theta}. 
\label{eq.Alsigma}
\end{eqnarray}
Here, we have assumed that the beam of light propagates perpendicular to the plane. We have used the paraxial approximation where the $z$-component of the vector potential can be neglected. We have used two quantities to characterize the light beam: $\sigma=\pm 1$ to represent circular polarization and $\kappa$, taking any integer value,to denote the orbital angular momentum. The amplitude profile of the light beam is taken to be $F(r)$. For example, in a Bessel beam, this profile takes the form of a Bessel function. 

Combining Eqs.~\ref{eq.deltapsi} and \ref{eq.Alsigma}, we find 
\begin{eqnarray}
\nonumber \delta \psi (\mathbf{r},\delta t) = i(\delta t)e^{-i\theta} \big[ &~&
 \tilde{F}(r) e^{i \{ \omega t - (\kappa + \sigma)\theta \} } \\
 &+& 
  \tilde{G}(r)e^{-i \{ \omega t - (\kappa + \sigma)\theta \} } 
\big].
\end{eqnarray}
We now specialize to the case of $\sigma=-1$. This choice is tied to the direction of the magnetic field or equivalently, to the sense of phase winding in the single vortex solution. With this choice, we find
\begin{eqnarray}
\nonumber \delta \psi (\mathbf{r},\delta t) = (\delta t) \big[ &~&
 \tilde{F}(r) e^{i \{ \omega t - \kappa \theta \} } \\
 &+& 
  \tilde{G}(r)e^{-i \{ \omega t + (2 -\kappa)\theta \} } 
\big].
\end{eqnarray}
Here, $\tilde{F} = -\frac{F(r)}{2}[I(r) + J(r)]$ and $\tilde{G} = -\frac{F(r)}{2}[I(r) - J(r)]$. Remarkably, we have arrived at precisely the same form that we have for normal modes. This can be seen by combining Eqs.~\ref{eq.etaform} and \ref{eq.uv_LL_expansion} above. In order to excite a normal mode, we require (a) $\omega = \epsilon$, i.e., the frequency of light should match the normal mode energy, and (b) $\kappa = m$, the orbital angular momentum must match the angular momentum quantum number of the desired normal mode. In particular, to excite a gyroscopic mode with $m=0$, we require purely circularly polarized light with $\kappa=0$. To exicte a breathing mode with $m=1$, we require twisted light with unit orbital angular momentum, i.e., $\kappa = 1$. When these conditions are met, the incident light will be absorbed. This allows for a sharp spectroscopy of the `atomic orbitals'. This is strongly reminiscent of the spectrosopic determination of electronic levels in the Hydrogen atom.

\subsection{Collective modes in the vortex lattice}

Excitations in vortex lattices have been extensively studied numerically, largely motivated by studies on ultracold atomic gases\cite{Mueller2003,Mizushima2004,Simula2013}. These studies typically consider a finite lattice with a trapping potential. As a consequence of working with finite systems, they do not consider the possibility of band formation. Our study provides new perspective that can extend these studies to infinite systems. 

As the first step, our discussion in Sec.~\ref{sec.VL} suggests how periodic boundary conditions can be implemented. This allows for explicit numerical solutions on an infinite vortex lattice. We note that pinning mechanisms can be explicitly included, e.g., in the form of an impurity potential and/or a Josephson coupling to a static substrate. The resulting collective modes can be fit to the form of Bloch's theorem given in Sec.~\ref{sec.bloch}. The Wannier function for each band can then be extracted and its physical character examined. The results of such a study can be understood using our tight binding approach.

Experimentally, Bragg scattering can be used to study collective excitations of vortex lattices in ultracold atomic gases\cite{Muniz2006}. A momentum-resolved study in a pinned vortex lattice can reveal the band structure of normal modes.

\section{Summary and discussion}
\label{sec.summary}
We present a study of collective modes about vortices and vortex lattices. Using the time-dependent Gross-Pitaevskii equation, we establish a notion of a fluctuation normal mode. 
In the case of an isolated vortex, we find low-lying modes corresponding to gyrotropic and breathing fluctuations. Given the recent interest in such modes in the context of skyrmions and `magnetic vortices', our results can motivate analogous studies in superfluids. 
In vortex lattices, we derive an analogue of Bloch's theorem that constrains the form of collective excitations. We develop a tight binding description where fluctuations can be viewed as hopping from one vortex to another. Although there is extensive literature on vortex lattices, a tight binding point of view has not been presented before. It presents a new route to study questions such as the stability of vortex lattices, transitions from square to triangular geometry, topological character in collective mode bands, etc. It also suggests tight binding as a promising paradigm to understand collective excitations in other emergent mesoscopic crystals, e.g., in skyrmion lattices.

In the first part of our study, we present a detailed study of collective modes around an isolated vortex. This strongly resonates with earlier studies on trapped ultracold atomic gases\cite{Svidzinsky2000,Svidzinsky2000b,Svidzinsky2001}. Numerical studies have identified low-lying excitations about a vortex as breathing modes and dipole modes\cite{Dodd1997}. The latter are precisely the gyrotropic modes discussed above. Gyrotropic modes have also been studied in a different guise -- as the precession of a vortex that is positioned off-centre in a cylindrical condensate\cite{Svidzinsky2000b,McGee2001}.
Our results in Sec.~\ref{sec.isolated} directly correspond to the excitations about a single vortex in a rotating Bose condensate. The trap potential can be viewed as providing the pinning potential.

The second part of our study discusses collective modes in a vortex lattice. This is a topic with a long history that stretches over decades. Earlier studies have largely taken a hydrodynamics-based approach, modelling vortex lattice fluctuations as coarse-grained deformations in a two-dimensional crystal. This point of view was shown to lead to Tkachenko modes, low-lying long-wavelength modes involving elliptical motion of vortices about their equilibrium positions\cite{Tkachenko1966,Fetter1967,Sonin1987,Reijnders2005}. These modes have been experimentally studied in ultracold atomic gases by perturbing a vortex lattice in various ways\cite{Engels2002,Coddington2003}. 
In this article, we have used a tight binding approach that is arguably more microscopic. It is also more general as it allows for a larger space of excitations. For example, our approach naturally allows for breathing modes where each vortex remains stationary, centred at its equilibrium position. Such excitations cannot be found in a hydrodynamic approach that is based on displacements of vortex centres.
The tight binding approach can provide useful insight into numerical collective mode spectra obtained by solving the time-dependent Gross-Pitaevskii equation\cite{Mueller2003,Mizushima2004,Simula2013}.

Our results can be contrasted with earlier studies that use the lowest Landau level approximation. 
This approach is justified for large magnetic fields, where the order parameter configuration can be written as an expansion in the lowest Landau level\cite{Thouless1975,Rosenstein2010}. Such a function (one within the lowest Landau level) is completely determined by the positions of its zeroes. Its energy depends purely on vortex positions. This suggests a view of the vortex lattice as a generalized ball-spring model with the energy given by a pairwise interaction potential. This approach has been used to study the melting of vortex lattices\cite{Herbut1994,Herbut1995,Menon1996,Li2004}. The lowest Landau level approach has also been used to understand collective modes about an ordered vortex lattice phase\cite{Matveenko2011,Yoshino2019}. These studies find gapless excitations that can be viewed as Tkachenko modes. Our tight binding scheme is more general in principle, as it is 
does not assume a strong magnetic field. However, it does not yield gapless excitations for generic parameters.  
This shortcoming originates at the level of a single vortex, as discussed in Sec.~\ref{ssec.landau}. Nevertheless,  our  approach is suitable for pinned vortex lattices. It brings out several new aspects that can be tested in experiments and numerical studies including band formation and mixing between breathing and gyrotropic modes.

Our study is based on the time-dependent Gross-Pitaevskii equation, well known to be a good description for ultracold atomic gases. However, our results may also have limited applicability to superconductors. Typically, dynamics in superconductors is modelled using the time-dependent Landau Ginzburg equation. It differs from the Gross-Pitaevskii equation in that the time derivative term does not come with a complex `i'\cite{Winiecki2001}. Physically, this represents dissipation with any deviation from the equilibrium solution decaying in time. In certain situations, it is conceivable that dissipation may not play a dominant role in a superconductor. Indeed, Ref.~\onlinecite{Barybin2011} has suggested a description that includes both disspation as well as coherent dynamics. This may result in coherent collective modes, but with strong decay due to damping. We believe this may be accessible in spectroscopic  measurements in suitable superconducting materials, as described in Sec.~\ref{ssec.expt_isolated} above.

\acknowledgments{
We thank Igor Herbut for useful comments. BBA thanks Prashanth Raman and Sourav Ballav for illuminating discussions. 
}

\renewcommand{\theequation}{A\arabic{equation}}
\renewcommand{\thefigure}{A\arabic{figure}}
\setcounter{equation}{0}
\setcounter{figure}{0}
\appendix
\section{Isolated vortex spectrum from the Landau level expansion}
\label{app.spectrum}

In this appendix, we discuss the normal modes about an isolated vortex. We seek to solve the eigenvalue problem posed in Eq.~\ref{eq.uveqs} with $\Psi_0(\mathbf{r})\equiv \Psi_{vortex}(\mathbf{r}) = \Delta(r) e^{-i\theta}$. As described in Eq.~\ref{eq.uv_LL_expansion}, we approach the problem by performing an expansion in Landau levels. The Landau level wavefunctions are given by 
\bea
\nonumber \psi^{LL}_{n,m}(\mathbf{r}) &=&  f^{LL}_{n,m}(r)e^{im\theta}; \\
\nonumber f^{LL}_{n,m}(r)
&=&\frac{(-1)^{n}}{\sqrt{2\pi \ell_{B}^{2}}}\sqrt{\frac{n!}{2^{m}(n+m)!}}\bigg(\frac{r}{\ell_{B}}\bigg)^{m}e^{-\frac{r^{2}}{4\ell_{B}^{2}}} \\
&\times&
L^{m}_{n}\bigg(\frac{r^{2}}{\ell_{B}^{2}}\bigg),
\eea
where $L^{m}_{n}(.)$ represents the associated Laguerre polynomial. In the following calculations, we work in units where $\ell_B$ is unity. As the Landau levels are stationary states in the problem of a free particle in a magnetic field, they satisfy
\bea
\frac{\big( -i\hbar\boldsymbol{\nabla} -2e\textbf{A} (\mathbf{r})\big)^{2} }{2m^*} \psi^{LL}_{n,m}(\mathbf{r}) = \hbar \omega_c (n+ \frac{1}{2}) \psi^{LL}_{n,m}(\mathbf{r}),~~~
\eea
where $\omega_c$ is the cyclotron frequency.
In the calculations below, we use two orthogonality relations arising from overlaps of Landau levels,
\bea
\int d^2 r  \big(\psi^{LL}_{n',m'}(\mathbf{r})\big)^*  \psi^{LL}_{n,m}(\mathbf{r}) = \delta_{n,n'} \delta_{m,m'},\\
 \int d^2 r  \big(\psi^{LL}_{n',m'}(\mathbf{r})\big)^* \big\{ \phi(r)e^{i\tilde{m}\theta} \big\}\psi^{LL}_{n,m}(\mathbf{r}) \propto \delta_{m+\tilde{m},m'}. 
\eea 
The first is a straightforward orthonormality relation. In the second, we have the matrix element of an arbitrary function with a definite angular momentum, $\phi(r)e^{i\tilde{m}\theta}$. Here, the integral over the angular coordinate, $\theta$, enforces a delta function on the $m$'s.

We now consider Eq.~\ref{eq.uveqs} with $u(\mathbf{r})$ and $v(\mathbf{r})$ expanded in the basis of Landau levels.
Using the orthogonality relations, we find that states with angular momentum $m$ are coupled to those with angular momentum $2-m$. This leads to a matrix equation,
\bea
\left(
\begin{array}{cc}
-A_{m} & B_{m,2-m} \\
-B_{m,2-m} & A_{2-m}
\end{array}\right) 
\left(\begin{array}{c}
U_{m} \\ 
V_{2-m}
\end{array}\right) = \epsilon \left(\begin{array}{c}
U_{m} \\ 
V_{2-m}
\end{array}\right).
\label{eq.eig_eqn}
\eea
Here, $U_m = (u_{0,m}, u_{1,m}, \ldots)^T$ and $V_{2-m} = (v_{n_{min},2-m}, v_{n_{min}+1,2-m}, \ldots)^T$. For the $v$'s, the lowest value of the $n$ index is $n_{min} = \mathrm{max}[m-2,0]$. The entries $A_m$, $B_{m,2-m}$, etc. represent matrices. Their entries are given by 
\bea
\nonumber A_{m}(n,n') &=& [\hbar \omega_c (n+1/2) + a ]\delta_{n,n'} + 4\pi b I_m^{n',n},\\
B_{m,2-m}(n,n') &=& 2\pi b J_{m}^{n',n},
\label{eq.ABmatrices}
\eea 
where
\bea
I_{m}^{n',n} = \int_0^\infty r dr f_{n',m}^{LL} (r) \Delta^2 (r) f_{n,m}^{LL} (r),\\
J_m^{n',n} = \int_0^\infty r dr f_{n',2-m}^{LL} (r) \Delta^2 (r) f_{n,m}^{LL} (r),
\eea
where $\Delta(r)\approx \Delta_0 \tanh(\nu r/\xi)$ is the amplitude profile of an isolated vortex. 

We have arrived at an eigenvalue equation in Eq.~\ref{eq.eig_eqn}. We take the energy to be in units of $a$, the parameter in the Gross-Pitaevskii equation. To enforce this, we divide all elements of the matrix in  Eq.~\ref{eq.eig_eqn} by $a$. In the diagonal entries, we obtain a contribution proportional to $\hbar\omega_{c}/a$. We interpret this ratio as the ratio of two length scales as follows. The first is the magnetic length, defined as $\ell_B=\sqrt{\hbar/2eB}$. The second is the correlation length given by $\xi = \sqrt{\hbar^2/2m^{*}\vert a \vert}$. This is the length scale at which the order parameter recovers at the system boundary\cite{Annett_book}. Here, we assume that the same correlation length enters in the vortex profile. We now express $\frac{\hbar \omega_{c}}{2 \vert a \vert} = \frac{\hbar (2e B) (2m^{*} \xi^{2})}{2m^{*} \hbar^{2}} = \frac{2eB}{\hbar}\xi^{2} = \xi^{2}/\ell_{B}^{2}$. We treat this ratio of length scales as an input parameter that depends on the system, temperature, field, etc.

\begin{figure}\label{20.cutoff.spectrum}
\includegraphics[width=\columnwidth]{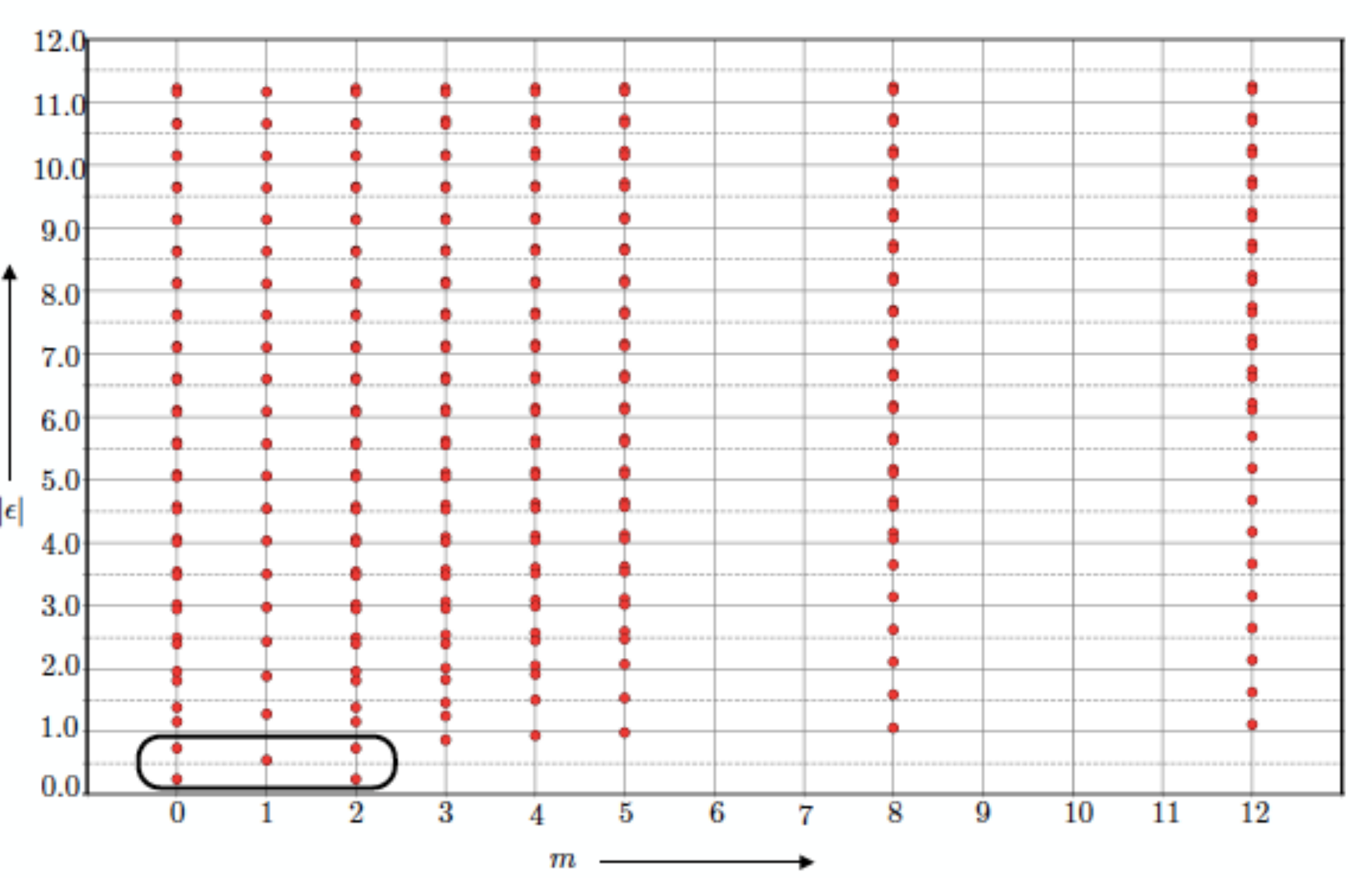}
\caption{Normal mode energies for a large choice of the cutoff with $n_{c}=20$. The lowest states occur in the $m=0$, $m=1$ and $m=2$ sectors, shown encircled.}
\label{fig.spec20}
\end{figure}

\begin{figure}\label{stability}
\includegraphics[width=\columnwidth]{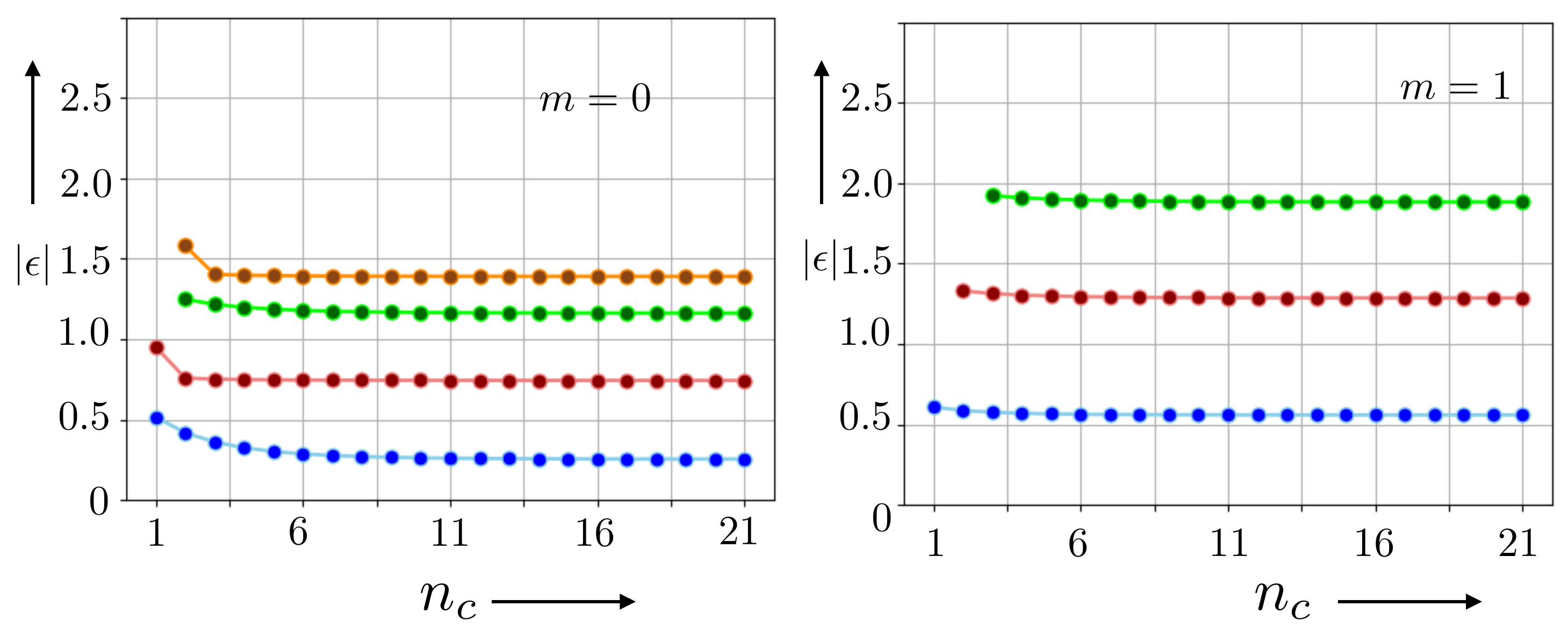}
\caption{Stability of the Eigenvalues as a function of the cutoff $n_{c}$. The left panel shows the energies in the $m=0$ sector. The right panel shows the energies for $m=1$. The first few levels are plotted in both the panels.}
\label{fig.spec_nc}
\end{figure}

The eigenvalue equation in Eq.~\ref{eq.eig_eqn} involves a matrix that is infinite-dimensional. The $U_m$ and $V_{2-m}$ arrays each comprise an infinite number of terms, one for each value of the $n$ index. 
To study Eq.~\ref{eq.uv_eigeqn} in a systematic manner, we introduce a cut-off in $n$, keeping $n=0,1,\ldots,(n_c-1)$ in the $u$'s. In the $v$'s, we keep $n=n_{min},\ldots,(n_c-1)$. If $n_{min}> (n_c-1)$, we do not have any $v$'s in Eq.~\ref{eq.uv_eigeqn}. For $m=0,1$, or $2$, this scheme results in a $2n_c\times 2n_c$ matrix that yields $2n_c$ eigenvalues. For higher values of $m$, we obtain a smaller matrix of size $\mathrm{max}[2n_c - n_{min},n_c]$. 

Fig.~\ref{fig.spec20} shows the resulting eigenvalue spectrum vs. $m$, for a large choice of the cutoff. We deduce two important features from this figure: (i) For each $m$, the eigenenergies form a discrete spectrum with well separated eigenvalues.
(ii) The energies generally increase with increasing $m$, saturating at a certain threshold. As we are interested in excitations with the lowest energy, we may restrict our attention to $m=0,1,2$ below. (iii) The lowest few energies do not change strongly with $n_c$. This can be seen from Fig. \ref{fig.spec_nc} that shows the spectrum for $m=0,1,2$ as a function of the cutoff, $n_c$. Crucially, the eignenergies do not change significantly with $n_c$.  It follows that we can obtain a good description of low energy modes by choosing the smallest cut-off, $n_c=1$. 
 
Based on these arguments, we see that the lowest lying excitations can be found as eigenvalues of a $2\times 2$ matrix, as given in Eq.~\ref{eq.uv_eigeqn} of the main text. The explicit entries of this matrix can be found by setting $n=n'=0$ in Eq.~\ref{eq.ABmatrices}. We present numerical results for the eigenvalues and eigenvectors in Tab.~\ref{tab.isolated}. These results have been obtained for $\xi/\ell_B = \frac{1}{2}$. For example, the first row in Tab.~\ref{tab.isolated} corresponds to the eigenvalue $0.516095a$, where $a$ is the Gross-Pitaevskii parameter. This is a negative quantity as $a < 0$ in the ordered phase. In accordance with the normalization condition below Eq.~\ref{eq.uveqs} in the main text, this normal mode satisfies $\vert u_{m} \vert^{2} - \vert v_{2-m} \vert^{2} = 1 $.

\begin{table}
\begin{tabular}{|c||c|c|c|}
\hline
$m$ & $\epsilon$ & $u_m$ & $v_{2-m}$ \\
\hline
0 & 0.516095 & 1.082382 & 0.414186 \\
\hline
0 & -0.952944 & 0.414186 & 1.082382 \\
\hline
1 & 0.608896 & 1.189009 & 0.643229 \\
\hline
1 & -0.608896 & 0.643229 & 1.189009 \\
\hline
\end{tabular}
\caption{Low energy excitations about an isolated vortex. These solutions have been found for $\xi/\ell_B=1/2$ with the cutoff chosen to be $n_c=1$. The eigenvalues $\epsilon$ are given in units of $a$, the Gross-Pitaevskii parameter.}
\label{tab.isolated}
\end{table}

\section{Effect of pinning}
\label{app.pinning}
\begin{figure}
\includegraphics[width=\columnwidth]{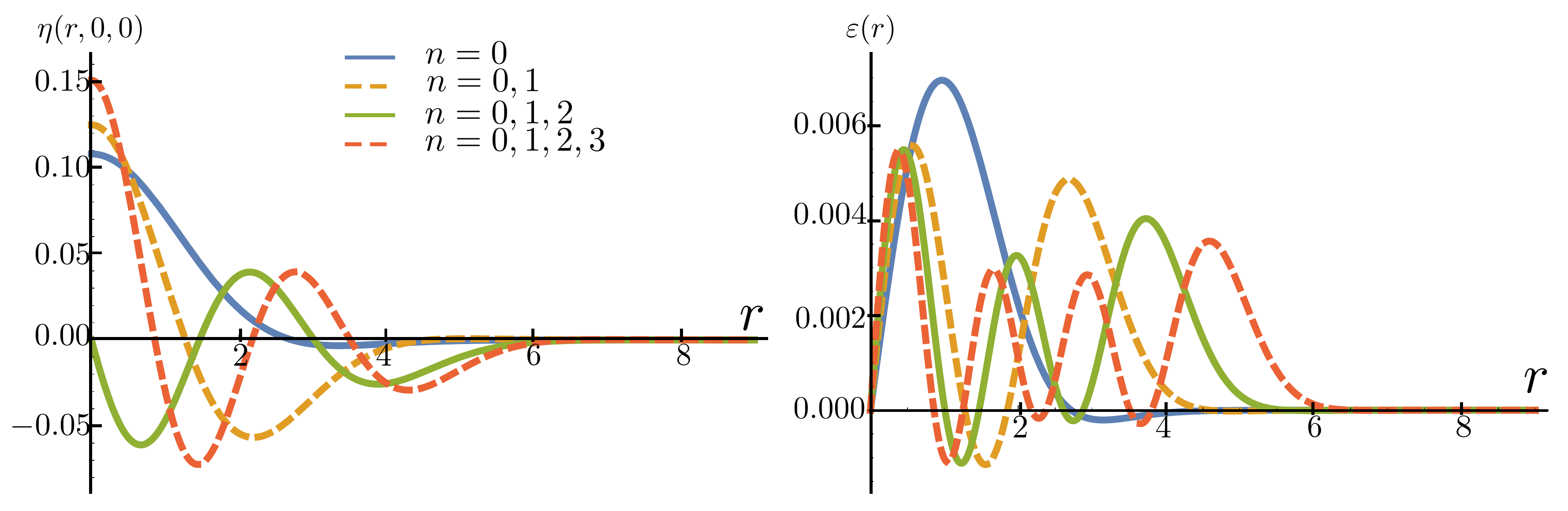}
\caption{Profile of the mode with the lowest energy in the $m=0$ sector for different choices of the Landau level cutoff. Left: Profile at time $t=0$ and along $\theta=0$. Right: The energy density in the mode. The mode becomes longer-ranged as more Landau levels are included.} 
\label{fig.ground}
\end{figure}
\begin{figure}
\includegraphics[width=\columnwidth]{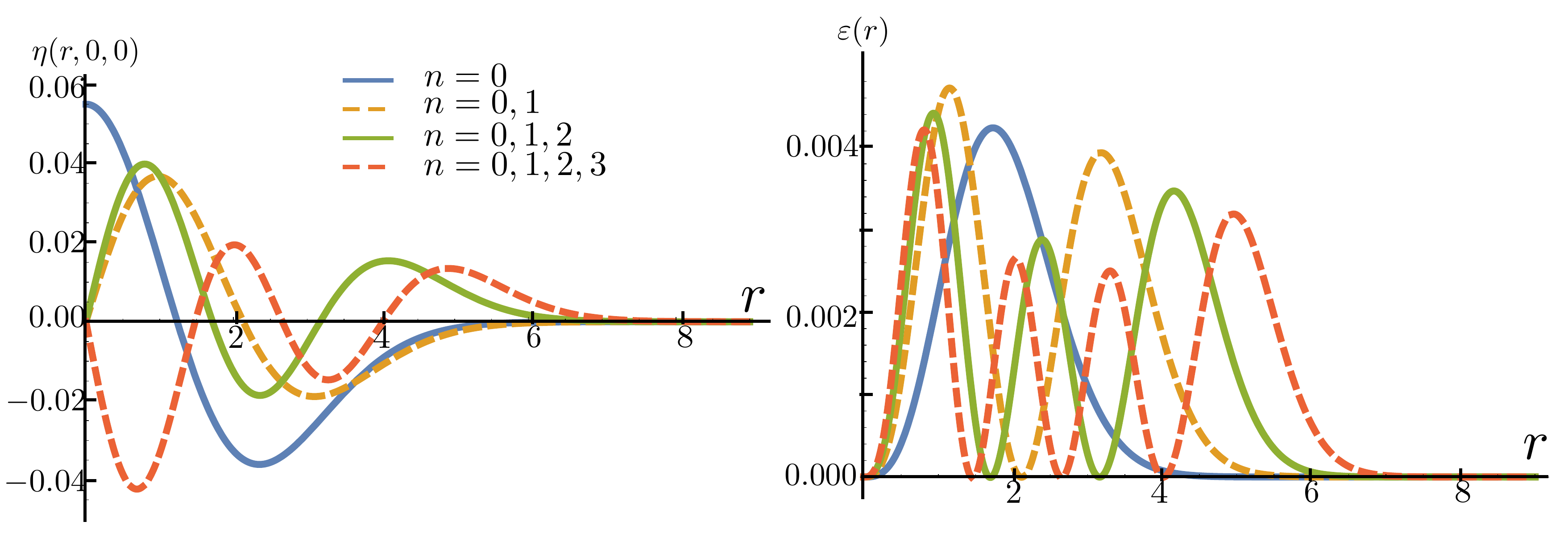}
\caption{Profile of the mode with the lowest energy in the $m=1$ sector with different Landau level cutoffs. Left: Profile at time $t=0$ and along $\theta=0$. Right: The energy density in the mode. The mode becomes longer-ranged as more Landau levels are included.}
\label{fig.first}
\end{figure}

In Sec.~\ref{ssec.landau} of the main text, we have asserted that the Landau level cutoff can be viewed as encoding the strength of pinning potentials. We now support this assertion by tracking the change in the normal mode wavefunction upon increasing $n_c$. 
As $n_{c}$ increases, the low energy modes become progressively longer-ranged. This is shown in Fig.~\ref{fig.ground} for the lowest energy mode in the $m=0$ sector and in Fig.~\ref{fig.first} for the $m=1$ sector. The plots show the profile of the wavefunction, at time $t=0$ along the ray defined by $\theta=0$. The modes themselves as well as their energy content expand in radius with increasing $n_c$. This is precisely the behaviour that one would expect to see when the strength of pinning potentials decreases. 

The behaviour seen in Figs.~\ref{fig.ground} and \ref{fig.first} follows from the well-known properties of Landau level wavefunctions. As the $n$ quantum number increases, they become longer-ranged. As our collective modes are composed of Landau level wavefunctions, they naturally show the same behaviour with increasing $n_c$. 

\section{Vortex lattices under translations}
\label{app.VL_trans}
In the main text, we have established the effect of lattice translations on a vortex lattice solution. In Eq.~\ref{eq.psi_trans_constraints}, we find that the solution accrues a phase originating from the gauge structure, as well as an additional phase. We now consider two translations by $\mathbf{R}_{p,q}$ and $\mathbf{R}_{p',q'}$. Applying these translations in sequence, Eq.~\ref{eq.psi_trans_constraints} leads to
\bea
\nonumber \Psi_{VL}(\mathbf{r}) &=& e^{-i\theta_{p,q}} e^{2ie\lambda_{p,q}(\mathbf{r})/\hbar} \Psi_{VL}(\mathbf{r} - \mathbf{R}_{p,q}) \\
\nonumber &=&e^{-i\theta_{p,q}} e^{2ie\lambda_{p,q}(\mathbf{r})/\hbar}  
 e^{i\theta_{p',q'}}e^{2ie\lambda_{p',q'}(\mathbf{r}-\mathbf{R}_{p,q})/\hbar} \\
&~& \times \Psi_{VL}(\mathbf{r} - \mathbf{R}_{p,q}-\mathbf{R}_{p',q'}).~~~
\eea
We now compare this with a direct translation by $\mathbf{R}_{p+p',q+q'}$, giving
\bea
\nonumber \Psi_{VL}(\mathbf{r}) &=& e^{-i\theta_{p+p',q+q'}} e^{2ie\lambda_{p+p',q+q'}(\mathbf{r})/\hbar}\times \\ 
&~& \Psi_{VL}(\mathbf{r} - \mathbf{R}_{p+p',q+q'}). 
\eea
Comparing these two relations, we obtain
\bea
\nonumber &~& \theta_{p,q} + \theta_{p',q'} - \theta_{p+p',q+q'} \equiv \frac{2e}{\hbar} \times \\
\nonumber &~&   \big\{
 \lambda_{p,q}(\mathbf{r}) + \lambda_{p',q'}(\mathbf{r}-\mathbf{R}_{p,q})
-\lambda_{p+p',q+q'}(\mathbf{r}) 
\big\}\\
&~&  =  \frac{2e}{\hbar} \frac{B}{2} \{ \hat{z} \cdot (\mathbf{R}_{p,q} \times \mathbf{R}_{p',q'})\}.
\label{eq.theta_relations}
\eea
The `$\equiv$' sign signifies that this is an equality between angles, i.e., the two sides may differ by an integer multiple of $2\pi$. 
In the last step, we have used the definition of $\lambda_{p,q}(\mathbf{r})$, given below Eq.~\ref{eq.Atrans} in the main text. 

We now use Eq.~\ref{eq.theta_relations} to determine the allowed values of $\theta_{p,q}$. We first consider translations purely along the first primitive lattice vector, $\hat{a}$. That is, we consider $\mathcal{T}_{p,0}$ where $p$ can take any value. Setting $q=q'=0$ in Eq.~\ref{eq.theta_relations}, we find 
\bea
 \theta_{p,0} + \theta_{p',0} = \theta_{p+p',0}. 
\eea 
This is a powerful relation, showing that $\theta_{p,0}$ scales linearly with $p$. We satisfy this relation by assigning $\theta_{p,0} \equiv p \theta_a$, where $\theta_a$ is an arbitrary constant. Similarly, considering translations purely along the other primitive lattice vector, $\hat{b}$, we have $\theta_{0,q} \equiv q \theta_b$, where $\theta_b$ is another arbitrary constant. 

We now consider lattice translations that have non-zero projections along both primitive lattice vectors. We first simplify Eq.~\ref{eq.theta_relations},
\bea
 \theta_{p,q} + \theta_{p',q'} - \theta_{p+p',q+q'} \equiv  \frac{e}{\hbar} B (p q' - qp') (\hat{z}\cdot \hat{a} \times \hat{b}).~~~
\eea
We now identify $\{B (\hat{z}\cdot \hat{a} \times \hat{b})\}$ as the magnetic flux through a single plaquette of the vortex lattice. As each plaquette carries a single flux quantum, we have $B (\hat{z}\cdot \hat{a} \times \hat{b}) = h/2e$. We obtain 
\bea
\theta_{p,q} + \theta_{p',q'} - \theta_{p+p',q+q'} \equiv  \pi (pq' - qp').
\label{eq.theta_mnpi}
\eea  
Considering $q=p'=0$, we immediately arrive at Eq.~\ref{eq.theta_mn} in the main text.

\section{Normal modes under translations}
\label{app.mode_trans}

In the main text, Eq.~\ref{eq.uvtrans_phase} encapsulates the effect of a lattice translation on a normal mode. We first discuss its first component, $u(\mathbf{r})$. Upon translation by $\mathbf{R}_{p,q}$, it picks up a gauge-derived phase as well as an additional phase, $\beta_{p,q}$. Comparing Eqs.~\ref{eq.uvtrans_phase} and \ref{eq.psi_trans_constraints}, we see that $u(\mathbf{r})$ and the vortex lattice solution, $\Psi_{VL}(\mathbf{r})$ pick up similar phases upon lattice translations. It follows that Eq.~\ref{eq.theta_relations} applies with $\beta_{p,q}$ in place of $\theta_{p,q}$,
\bea
\nonumber &~& \beta_{p,q} + \beta_{p',q'} - \beta_{p+p',q+q'}  \\
&~&  \equiv  \frac{2e}{\hbar} \frac{B}{2} \{ \hat{z} \cdot (\mathbf{R}_{p,q} \times \mathbf{R}_{p',q'})\}
 = \pi(pq'-qp').~~~~~
 \label{eq.beta1}
\eea
In the last step, we have used the arguments leading up to Eq.~\ref{eq.theta_mnpi}.  
Using the same arguments as in Appendix~\ref{app.VL_trans}, we arrive at an analogue of Eq.~\ref{eq.theta_mn} from the main text,  
\bea
\beta_{p,q} = p \beta_a + q \beta_b + pq \pi,
\eea
where $\beta_a$ and $\beta_b$ are arbitrary constants. We now reinterpret these constants as components of a vector $\mathbf{k}$, with $\beta_a \equiv \mathbf{k} \cdot \hat{a}$ and $\beta_b \equiv \mathbf{k} \cdot \hat{b}$. With these relations, we have $(p \beta_a + q \beta_b) = \mathbf{k} \cdot \mathbf{R}_{p,q}$, so that $\beta_{p,q} = \mathbf{k}\cdot \mathbf{R}_{p,q} + p q \pi$.
It is easily seen that $\mathbf{k}$ can be taken to lie within the first Brillouin zone. Starting from an arbitrary choice for $\mathbf{k}$, we can always write $\mathbf{k}  = \mathbf{k}' - \mathbf{K}$, where $\mathbf{k}'$ lies in the first Brillouin zone and $\mathbf{K}$ is a reciprocal lattice vector. We have $\mathbf{k}\cdot\mathbf{R}_{p,q} \equiv \mathbf{k}'\cdot\mathbf{R}_{p,q} $, as the reciprocal lattice vector contributes an integer multiple of $2\pi$ to the dot product. 

We now consider the second component of the normal mode wavefunction, $v(\mathbf{r})$. Comparing the relation for $v(\mathbf{r})$ with that for $u(\mathbf{r})$ in Eq.~\ref{eq.uvtrans_phase}, we see that $\beta_{p,q}$ must satisfy a relation analogous to Eq.~\ref{eq.beta1}, but with a minus sign,
\bea
\nonumber &~& \beta_{p,q} + \beta_{p',q'} - \beta_{p+p',q+q'} \\
&~&  = - \frac{2e}{\hbar} \frac{B}{2} \{ \hat{z} \cdot (\mathbf{R}_{p,q} \times \mathbf{R}_{p',q'})\} = -\pi (pq'-qp').~~~~~~
\eea
This relation is consistent with Eq.~\ref{eq.beta1} as the right hand sides differ by an integer multiple of $2\pi$. This leads to the same form for $\beta_{p,q}$ as the arguments from $u(\mathbf{r})$ above. With these arguments, we arrive at Eq.~\ref{eq.beta_mn} in the main text.

\section{Translational invariance of overlap integrals}
\label{app.overlap_trans}

In this section, we consider the overlap integrals defined in the context of the tight binding model. We show that they satisfy the invariance properties that are required for a consistent tight binding scheme. We first consider the on-site overlap integral given by $s_{j',j}(p',q')$, as defined in Eq.~\ref{eq.s_def} of the main text. Using the form of the atomic orbitals given in Eq.~\ref{eq.eta}, we find that $s_{j',j}(p',q')$ is composed of four separate contributions, one from each element in the $(\delta H)_{p',q'}$ matrix. We represent them as $c_{ij}$, where $(ij)$ picks one element in $(\delta H)_{p',q'}$,
\bea
\nonumber &~& s_{j',j}(p',q') = c_{11} + c_{12} + c_{21} + c_{22}. 
\eea
We first consider the diagonal term
\bea
\nonumber &~& c_{11}= \int d^2 r ~-2b ( \vert \Psi_{VL}(\mathbf{r}) \vert^2 - \vert \Psi_{vortex,p',q'}(\mathbf{r})\vert^2 ) \times\\
\nonumber  &~& ~~~~~~~~~~~~~~~\{ u_{j'}^* (\mathbf{r} - \mathbf{R}_{p',q'}) u_{j} (\mathbf{r} - \mathbf{R}_{p',q'}) \}.
\eea
To see that this term is independent of $(p',q')$, we shift the integration variable to $\mathbf{r}' = \mathbf{r} - \mathbf{R}_{p',q'}$,
\bea
\nonumber c_{11}= \int d^2 r' ~-2b ( \vert \Psi_{VL}(\mathbf{r}') \vert^2 - \vert \Psi_{vortex}(\mathbf{r}')\vert^2 ) u_{j'}^* (\mathbf{r}') u_{j} (\mathbf{r}' ).
\eea
We have used the invariance of the vortex lattice amplitude under lattice translations. We have also used the form of $ \Psi_{vortex,p',q'}(\mathbf{r})$ given in Eq.~\ref{eq.psi_neighbourhood}. This form clearly shows that the $c_{11}$ contribution is the same at all sites. A similar argument works with $c_{22}$.  

We now consider an off-diagonal contribution 
\bea
\nonumber  c_{12}&=& \int d^2 r ~b \big(\Psi_{VL}^2(\mathbf{r})  - \Psi_{vortex,p',q'}^2 (\mathbf{r}) \big)\times \\
\nonumber &~& e^{-4ie\lambda_{p',q'}(\mathbf{r})/\hbar} u_{j'}^* (\mathbf{r} - \mathbf{R}_{p',q'}) 
v_{j} (\mathbf{r} - \mathbf{R}_{p',q'}) .
\eea
Using the translation properties given in Eqs.~\ref{eq.psi_trans_constraints_b} and \ref{eq.uv_mn}, we see that $e^{-4ie\lambda_{p',q'}(\mathbf{r})/\hbar}$ removes the non-trivial phase that is attached to 
$\big(\Psi_{VL}^2(\mathbf{r})  - \Psi_{vortex,p',q'}^2 (\mathbf{r}) \big)$. We then simplify the integral by shifting coordinates using $\mathbf{r}' = \mathbf{r} - \mathbf{R}_{p',q'}$. This leads to
\bea
\nonumber  c_{12}= \int d^2 r' ~b \big(\Psi_{VL}^2(\mathbf{r}')  - \Psi_{vortex}^2 (\mathbf{r}') \big)  u_{j'}^* (\mathbf{r}') v_{j} (\mathbf{r}') .
\eea
This is manifestly independent of $(p',q')$. A similar argument holds for the $c_{21}$ contribution as well.

We have shown that the on-site overlap, $s_{j',j}(p',q')$, is site-independent. It takes the same value for every $(p',q')$. This satisfies one of the requirements for a consistent tight binding formulation. We next consider the inter-site overlaps. We show that $r_{j',j}(p',q',p,q)$ depends only on the separation $(p-p',q-q')$. Similar arguments will also apply to the $t_{j',j}(p',q',p,q)$ overlaps. 

We use the explicit expressions in Eqs.~\ref{eq.deltaH} and \ref{eq.eta} to expand the $r$ overlaps defined in Eq.~\ref{eq.r_def}. It contains four terms, each corresponding to one term in the $(\delta H)_{p',q'}$ matrix,
\bea
r_{j',j}(p',q',p,q) = d_{11}+ d_{12}+ d_{21}+d_{22}.
\eea
We consider the the diagonal term,
\bea
\nonumber  d_{11} &=& \int d^2 r ~-2b ( \vert \Psi_{VL}(\mathbf{r}) \vert^2 - \vert \Psi_{vortex,p',q'}(\mathbf{r})\vert^2 ) \times\\
\nonumber  \big\{ u_{j'}^* (\mathbf{r} &-& \mathbf{R}_{p',q'}) u_{j} (\mathbf{r} - \mathbf{R}_{p,q})  e^{-2ie\lambda_{p'-p,q'-q}(\mathbf{r})/\hbar} e^{i(pq-p'q')\pi} \big\}.
\eea
Here, we have used the fact that $\lambda_{p,q}(\mathbf{r})$ is a linear function in $(p,q)$, e.g., $\lambda_{p,q}(\mathbf{r})+\lambda_{p',q'}(\mathbf{r})= \lambda_{p+p',q+q'}(\mathbf{r})$. 

Upon shifting the integration variable using ${\mathbf{r}'} = (\mathbf{r}-\mathbf{R}_{p',q'})$, this term takes the form
\bea
\nonumber ( \vert \Psi_{VL}(\mathbf{r}') \vert^2 &-& \vert \Psi_{vortex}(\mathbf{r}')\vert^2 ) \big\{ u_{j'}^* (\mathbf{r}') u_{j} (\mathbf{r}' - \mathbf{R}_{p-p',q-q'}) \} \\
 &\times& e^{-2ie\lambda_{p'-p,q'-q}(\mathbf{r}'-\mathbf{R}_{p',q'})/\hbar} e^{i(pq-p'q')\pi}.
\eea
Here, we have used the invariance in the amplitude of $\Psi_{VL}(\mathbf{r})$ under lattice translations. We have also used the form of $ \Psi_{vortex,p',q'}(\mathbf{r})$ given in Eq.~\ref{eq.psi_neighbourhood}.
By plugging in the definition of $\lambda$, we reexpress the phases to arrive at 
\bea
\nonumber ( \vert \Psi_{VL}(\mathbf{r}') \vert^2 &-& \vert \Psi_{vortex}(\mathbf{r}')\vert^2 ) \big\{ u_{j'}^* (\mathbf{r}') u_{j} (\mathbf{r}' - \mathbf{R}_{p-p',q-q'}) \} \\
&\times&  e^{-2ie\lambda_{p'-p,q'-q}(\mathbf{r}')/\hbar} e^{i(p-p')(q-q')\pi}.
\eea
We have now arrived at a form that depends only on $(p-p',q-q')$, but not on $(p',q')$. Similar arguments hold for the $d_{22}$ term as well.

We now consider an off-diagonal term 
\bea
\nonumber d_{12} &=& \int d^2 r ~b 
\big(\Psi_{VL}^2(\mathbf{r})  - \Psi_{vortex,p',q'}^2 (\mathbf{r}) \big) \times\\
\nonumber  \big\{ u_{j'}^* (\mathbf{r} &-& \mathbf{R}_{p',q'}) v_{j} (\mathbf{r} - \mathbf{R}_{p,q})  e^{-2ie\lambda_{p'+p,q'+q}(\mathbf{r})/\hbar} e^{i(pq-p'q')\pi} \big\}.
\eea
We shift the integration variable to ${\mathbf{r}'} = (\mathbf{r}-\mathbf{R}_{p',q'})$. Using Eqs.~\ref{eq.psi_trans_constraints_b} and \ref{eq.uv_mn}, we have 
\bea
\nonumber \big(\Psi_{VL}^2(\mathbf{r})  &-& \Psi_{vortex,p',q'}^2 (\mathbf{r}) \big)  = \\
\big(\Psi_{VL}^2(\mathbf{r}')  &-& \Psi_{vortex}^2 (\mathbf{r}') \big)  e^{4ie\lambda_{p',q'}(\mathbf{r}' + \mathbf{R}_{p',q'})/\hbar}.
\eea
Using the definition of $\lambda$, after a few simple manipulations, we arrive at
\bea
\nonumber d_{12} &=& \int d^2 r' ~b 
\big(\Psi_{VL}^2(\mathbf{r}')  - \Psi_{vortex}^2 (\mathbf{r}') \big) \times\\
\nonumber   u_{j'}^* (\mathbf{r}') v_{j} (\mathbf{r}' &-& \mathbf{R}_{p-p',q-q'})  e^{2ie\lambda_{p'-p,q'-q}(\mathbf{r})/\hbar} e^{i(p-p')(q-q')\pi} .
\eea
This form solely depends on the bond displacement $(p-p',q-q')$ and not on $(p,q)$ or $(p',q')$. This satisfies translational invariance as required for the tight binding formulation. A similar argument holds for the $d_{21}$ term. We have shown that the $r$ overlaps depend only on the separation between sites. On the same lines, it can be seen that the $t$ overlaps also depend on the separation alone.

\section{Tight binding on the square lattice}
\label{app.tb_rotations}

We provide details about the tight binding calculation on the square lattice here. We use the results from Appendix~\ref{app.spectrum} for atomic orbitals. These were calculated for a certain value of the correlation length, by setting $\xi/\ell_B=\frac{1}{2}$. We set up the tight binding model using the four lowest energy states as atomic orbitals. 

We first describe the overlap integrals, starting with the on-site overlaps. We evaluate them as defined in Eq.~\ref{eq.s_def}. In Appendix~\ref{app.overlap_trans}, we have shown that the on-site overlap is site-independent, i.e., $s_{j',j}(p',q')\equiv s_{j',j}$. We now evaluate these overlaps, taking $(p',q')$ to be the origin for simplicity. The first on-site overlap, corresponding to $j=j'=1$, is given by
\begin{widetext}
\bea
\nonumber s_{1,1} & = & -2b |u_{1}|^2 \int d^2 r( \vert \Psi_{VL}(\mathbf{r} ) \vert^2 - \vert \Psi_{vortex,p',q'}(\mathbf{r})\vert^2 ) ~  \psi^{LL}_{0,0} (\mathbf{r} )\{ \psi^{LL}_{0,0} (\mathbf{r} ) \}^* \\
\nonumber &-&  2b |v_{1}|^2 ~ \int d^2 r ( \vert \Psi_{VL}(\mathbf{r} ) \vert^2 - \vert \Psi_{vortex,p',q'}(\mathbf{r} )\vert^2 ) ~ \{ \psi^{LL}_{2,2} (\mathbf{r} ) \}^* \psi^{LL}_{2,2} (\mathbf{r} ) \\
\nonumber & + & bu_{1}^*v_{1}^* \int d^2 r ~ \big(\Psi_{VL}^2(\mathbf{r})  - \Psi_{vortex}^2 (\mathbf{r} ) \big) ~   \psi^{LL}_{0,0} (\mathbf{r} )  \psi^{LL}_{0,2} (\mathbf{r} ) \\
 & + & bv_{1}u_{1}\int d^2 r ~ \big(\Psi_{VL}^{*2}(\mathbf{r})  - \Psi_{vortex}^{*2} (\mathbf{r}) \big) ~ \{ \psi^{LL}_{0,2} (\mathbf{r} ) \}^* \{ \psi^{LL}_{0,0} (\mathbf{r} )\}^* = 0.184799.
\eea
\end{widetext}
Here, we have assumed that the vortex lattice amplitude is given by the ersatz functional form given in Eq.~\ref{eq.sqVLform}. In the third and fourth lines, we also require the phase of $ \Psi_{VL}(\mathbf{r} )$. Here, we assume that the dominant contribution to the integrals comes from the near vicinity of the origin. In this region, the phase of the vortex lattice resembles that of an isolated vortex with $\Psi_{VL}(\mathbf{r} ) \approx \vert \Psi_{VL}(\mathbf{r} ) \vert e^{-i\theta}$. We use these arguments to evaluate this overlap and all other on-site overlaps. 
Their values are tabulated in Tab.~\ref{tab.svals}.

\begin{table}
\begin{tabular}{|c|c|c|c|c|}
\hline
\diagbox{$j$ }{$j'$} & 1 & 2 & 3 & 4 \\
\hline
1 & 0.184799 & 0.144795 & 0 & 0 \\
\hline
2 & 0.144795 & 0.576461 & 0 & 0 \\
\hline
3 & 0 & 0 & 0.441242 & 0.255718 \\
\hline
4 & 0 & 0 & 0.255718 & 0.441242 \\ 
\hline
\end{tabular}
\caption{On-site overlaps, $s_{j,j'}$ evaluated for the square lattice. }
\label{tab.svals}
\end{table}

We now present details regarding the inter-site overlaps. As discussed in the main text, we restrict ourselves to nearest neighbours. We first take up the $t$ overlaps. Their evaluation follows in a straightforward manner from the definition in Eq.~\ref{eq.t_def} using the results from Appendix~\ref{app.spectrum}. For example, on the horizontal bond that connects sites $(0,0)$ and $(1,0)$, we have
\begin{widetext}
\bea
\nonumber t_{1,1}((p^{\prime},q^{\prime} \equiv 0,0),( p,q \equiv 1,0))  &=& \vert u_{1} \vert^2 \int d^2 r ~ \psi^{LL}_{0,0} (\mathbf{r})  \big\{ \psi^{LL}_{0,0} (\mathbf{r} - \ell \hat{x}) \big\}^* ~  e^{\frac{i\pi}{\ell} \hat{z}.[\textbf{r} \times \hat{x}]} \\
&-& \vert v_{1}\vert^2 \int d^2 r ~ \big\{ \psi^{LL}_{0,2} (\mathbf{r} )\big\}^*  \psi^{LL}_{0,2} (\mathbf{r}-\ell \hat{x})  ~ e^{-\frac{i\pi}{\ell} \hat{z}.[\textbf{r} \times \hat{x}]}  \approx 0.20788.
\eea
\end{widetext}
The remaining $t$ overlaps on this bond can be calculated in similar fashion. Their values are tabulated in Tab.~\ref{tab.tvals}. As discussed in Appendix~\ref{app.overlap_trans}, these overlap values have translation symmetry. They take the same value for any bond that is connected by the same lattice vector, i.e., for any bond connecting $(p,q)$ with $(p,q+1)$.

We have given explicit values for the $t$ overlaps for a horizontal bond in Tab.~\ref{tab.tvals}. For the remaining three nearest neighbour bonds, the corresponding values can be found from symmetry arguments. Denoting the $t$ overlaps as $t_{j,j'}(p-p',q-q')$, we find 
\bea
\nonumber t_{j,j'}(0,1) = & t_{j,j'}(1,0) e^{-i \pi/ 2}, \\
\nonumber t_{j,j'}(0,-1) =  & t_{j,j'}(1,0) e^{-i3\pi/2 }, \\
\nonumber t_{j,j'}(-1,0) = & t_{j,j'}(1,0)e^{-i\pi}.
\eea 
These relations can be deduced from the transformation properties of Landau level basis states under a four-fold rotation. Similar transformation properties hold for all $t$ and $r$ overlap integrals on all bonds.

\begin{table}
\begin{tabular}{|c|c|c|c|c|}
\hline
\diagbox{$j$ }{$j'$}  & 1 & 2 & 3 & 4 \\
\hline
1 & 0.20788 & $0^{*}$ & $0^{*}$ & $0^{*}$ \\
\hline
2 & $0^{*}$ & -0.20788 & $0^{*}$ & $0^{*}$ \\
\hline
3 & $0^{*}$ & $0^{*}$ & 0.20788 & 0 \\
\hline
4 & $0^{*}$ & $0^{*}$ & 0 & -0.20788 \\
\hline
\end{tabular}
\caption{Nearest neighbour $t$ overlaps on the horizontal nearest neighbour bond. Values on other nearest neighbour bonds are related by symmetry. The values indicated by $0^{*}$ are smaller than $10^{-5}$. We have set them to zero in our calculations. The diagonal terms are equal in amplitude upto six significant digits.}
\label{tab.tvals}
\end{table}

We now discuss the inter-site $r$ overlaps on nearest neighbours. We use the ersatz form of the vortex lattice amplitude given in Eq.~\ref{eq.sqVLform} of the main text. We make a further simplifying assumption. We assume that the dominant contribution in the $r$-overlap comes from the vicinity of the reference site. In this region, we may take the vortex lattice solution to have a simple phase winding relation. That is, we assume $\Psi_{VL}(\mathbf{R}_{p',q'} + \delta r) \approx \vert \Psi_{VL} (\mathbf{R}_{p',q'} + \delta r)\vert e^{i\theta_{p',q'}}$. Here, $\theta_{p',q'}$ is the polar angle about the point $\mathbf{R}_{p',q'}$.

To give a concrete illustration, we take the $r$ overlap corresponding to $j=j'=1$ on the horizontal bond connecting sites $(0,0)$ and $(1,0)$,
\begin{widetext}
\bea
\nonumber r_{1,1}((p^{\prime},q^{\prime} \equiv 0,0),( p,q \equiv 1,0))  & = & -2b |u_{1}|^{2} \int d^2 r ~ ( \vert \Psi_{VL}(\mathbf{r}) \vert^2 - \vert \Psi_{vortex}(\mathbf{r})\vert^2 ) 
 \psi^{LL}_{0,0} (\mathbf{r})  \{ \psi^{LL}_{0,0} (\mathbf{r} - \ell \hat{x}) \big\}^* e^{\frac{i\pi}{\ell} \hat{z}.[\textbf{r} \times \hat{x}]}\\
\nonumber &~& -2b |v_{1}|^{2} \int d^2 r ~ ( \vert \Psi_{VL}(\mathbf{r}) \vert^2 - \vert \Psi_{vortex}(\mathbf{r})\vert^2 ) 
\{ \psi^{LL}_{0,2} (\mathbf{r})\}^*  \psi^{LL}_{0,2} (\mathbf{r} - \ell \hat{x})  e^{-\frac{i\pi}{\ell} \hat{z}.[\textbf{r} \times \hat{x}]} \\
\nonumber &~& + ~ b u_{1}^*v_{1}^*\int d^2 r  
\big(\Psi_{VL}^2(\mathbf{r})  - \Psi_{vortex}^2 (\mathbf{r}) \big)   \psi^{LL}_{0,0} (\mathbf{r}) \psi^{LL}_{0,2} (\mathbf{r} - \ell \hat{x})  e^{-\frac{i\pi}{\ell} \hat{z}.[\textbf{r} \times \hat{x}]} \\
\nonumber &~& + ~ b v_{1}u_{1}\int d^2 r  
\big(\Psi_{VL}^{*2}(\mathbf{r})  - \Psi_{vortex}^{*2} (\mathbf{r}) \big) \{ \psi^{LL}_{0,2} (\mathbf{r}) \}^* \{ \psi^{LL}_{0,0} (\mathbf{r} - \ell \hat{x}) \}^*  e^{\frac{i\pi}{\ell} \hat{z}.[\textbf{r} \times \hat{x}]}  \approx 0.058659.
\eea
\end{widetext}
The remaining $r$ overlaps on this bond can be calculated in similar fashion. Their numerical values are tabulated in Tab.~\ref{tab.rvals}. As shown in Appendix~\ref{app.overlap_trans}, we obtain the same values on all bonds that are obtained by lattice translations. The values on other nearest neighbour bonds are dictated by symmetry. Representing the $r$ overlaps on a bond connecting sites $(p,q)$ and $(p',q')$ as $r_{j,j'}(p-p',q-q')$, we have 
\bea
\nonumber r_{j,j'}(0,1) = & r_{j,j'}(1,0) e^{-i\pi/2} \\
\nonumber r_{j,j'}(0,-1) =  & r_{j,j'}(1,0) e^{-i3\pi/2} \\
\nonumber r_{j,j'}(-1,0) = & r_{j,j'}(1,0)e^{-i\pi}.
\eea 
These relations can be deduced from transformation properties of Landau level states under fourfold rotations.

\begin{table}
\begin{tabular}{|c|c|c|c|c|}
\hline
\diagbox{$j$ }{$j'$} & 1 & 2 & 3 & 4 \\
\hline
1 & 0.058659 & 0.026673 & 0.034139 & 0.017851 \\
\hline
2 & 0.006460 & 0.090857 & 0.015871 & -0.1399 \\
\hline
3 & 0.07200 & 0.032275 & -0.049461 & -0.122866 \\
\hline
4 & 0.032275 & 0.092993 & -0.122866 & -0.013801 \\
\hline
\end{tabular}
\caption{Nearest neighbour $r$ overlaps. The values given here are for a horizontal nearest neighbour bond. Values on other nearest neighbours are related by symmetry. }
\label{tab.rvals}
\end{table}

Having determined the overlaps, we set up the tight binding matrix, $M_{ij}(\mathbf{k},\varepsilon_\mathbf{k})$. Using the symmetry relations between nearest neighbour bonds, the matrix takes the form,

\begin{widetext}
\footnotesize
\bea
\nonumber &~& M_{ij}(\mathbf{k},\varepsilon_\mathbf{k})  \\
\nonumber &=& T_{i,j}(\mathbf{k},\varepsilon_\mathbf{k}) + R_{i,j} + S_{i,j}\\
\nonumber &=& \left(
\begin{array}{c|c|c|c}
(\epsilon_{1} - \varepsilon_{\textbf{k}})[t_{1,1}(1,0)f(k_{x},k_{y}) + 1] & (\epsilon_{1} - \varepsilon_{\textbf{k}})t_{1,2}(1,0)f(k_{x},k_{y})  & (\epsilon_{1} - \varepsilon_{\textbf{k}})t_{1,3}(1,0)g(k_{x},k_{y})  & (\epsilon_{1} - \varepsilon_{\textbf{k}})t_{1,4}(1,0)g(k_{x},k_{y})  \\ \hline
(\epsilon_{1} - \varepsilon_{\textbf{k}})t^{*}_{1,2}(1,0)f^{*}(k_{x},k_{y})  & (\epsilon_{2} - \varepsilon_{\textbf{k}})[t_{2,2}(1,0)f(k_{x},k_{y}) - 1] & (\epsilon_{2} - \varepsilon_{\textbf{k}})t_{2,3}(1,0)g(k_{x},k_{y})  & (\epsilon_{2} - \varepsilon_{\textbf{k}})t_{2,4}(1,0)g(k_{x},k_{y})  \\ \hline
(\epsilon_{1} - \varepsilon_{\textbf{k}})t^{*}_{1,3}(1,0)g^{*}(k_{x},k_{y})  & (\epsilon_{2} - \varepsilon_{\textbf{k}})t^{*}_{2,3}(1,0)g^{*}(k_{x},k_{y})  & (\epsilon_{3} - \varepsilon_{\textbf{k}})[t_{3,3}(1,0)f(k_{x},k_{y}) + 1] & (\epsilon_{3} - \varepsilon_{\textbf{k}})t_{3,4}(1,0)f(k_{x},k_{y})  \\ \hline
(\epsilon_{1} - \varepsilon_{\textbf{k}})t^{*}_{1,4}(1,0)g^{*}(k_{x},k_{y})  & (\epsilon_{2} - \varepsilon_{\textbf{k}})t^{*}_{2,4}(1,0)g^{*}(k_{x},k_{y})  & (\epsilon_{3} - \varepsilon_{\textbf{k}})t^{*}_{3,4}(1,0)f^{*}(k_{x},k_{y})  & (\epsilon_{4} - \varepsilon_{\textbf{k}})[t_{4,4}(1,0)f(k_{x},k_{y}) - 1]
\end{array}
\right) \\
\normalsize
\nonumber &+& \left(
\begin{array}{c|c|c|c}
r_{1,1}(1,0)f(k_{x},k_{y}) & r_{1,2}(1,0)f(k_{x},k_{y}) & r_{1,3}(1,0)g(k_{x},k_{y}) & r_{1,4}(1,0)g(k_{x},k_{y}) \\ \hline
r^{*}_{1,2}(1,0)f^{*}(k_{x},k_{y}) & r_{2,2}(1,0)f(k_{x},k_{y}) & r_{2,3}(1,0)g(k_{x},k_{y}) & r_{2,4}(1,0)g(k_{x},k_{y}) \\ \hline
r^{*}_{1,3}(1,0)g*(k_{x},k_{y}) & r^{*}_{2,3}(1,0)g^{*}(k_{x},k_{y}) & r_{3,3}(1,0)f(k_{x},k_{y}) & r_{3,4}(1,0)f(k_{x},k_{y}) \\ \hline
r^{*}_{1,4}(1,0)g^{*}(k_{x},k_{y}) & r^{*}_{2,4}(1,0)g^{*}(k_{x},k_{y}) & r^{*}_{3,4}(1,0)f(k_{x},k_{y}) & r_{4,4}(1,0)f(k_{x},k_{y})
\end{array}
\right) + \left(
\begin{array}{c|c|c|c}
s_{1,1} & s_{1,2} & s_{1,3} & s_{1,4} \\ \hline
s^{*}_{1,2} & s_{2,2}& s_{2,3} & s_{2,4} \\ \hline
s^{*}_{1,3} & s^{*}_{2,3} & s_{3,3} & s_{3,4} \\ \hline
s^{*}_{1,4} & s^{*}_{2,4} & s^{*}_{3,4} & s_{4,4}
\end{array}
\right),
\eea
\end{widetext}
where we have used, 
\bea
\nonumber f(k_{x},k_{y}) &= 2[\cos(k_{x}) + \cos(k_{y})],\\
\nonumber g(k_{x},k_{y}) &= 2[i\sin(k_{x}) - \sin(k_{y})].
\eea
The normal mode energies, $\varepsilon_\mathbf{k}$, are obtained by solving $\mathrm{Det}(M_{ij}  (\mathbf{k},\varepsilon_\mathbf{k}) = 0$.

\bibliographystyle{apsrev4-1} 
\bibliography{vortex_tb}
\end{document}